\documentclass[a4paper,11pt,pdftex]{article}
\pdfoutput=1

\usepackage{jheppub}

\usepackage[utf8]{inputenc}
\usepackage{amsfonts}
\usepackage{amsbsy}
\usepackage{amssymb}
\usepackage{url}
\usepackage{enumerate}
\usepackage{caption}
\usepackage{hhline}

\usepackage{graphicx,xcolor}
\usepackage{amssymb,amsmath}
\usepackage{slashed}
\usepackage{hyperref}
\usepackage{braket}
\usepackage{simplewick}
\usepackage{ascmac}
\usepackage{latexsym}
\usepackage{pifont}          
\usepackage{bm}
\usepackage{comment}

\usepackage{subfigure,physics}
\usepackage{mathtools,autobreak}

\newcommand{\n}{\nonumber}

\newcommand{\mr}[1]{\mathrm{#1}}

\newcommand{\f}[2]{\frac{#1}{#2}}
\newcommand{\s}[1]{\slashed{#1}}

\newcommand{\yh}[1]{\textcolor{magenta}{#1}}

\makeatletter
\gdef\@fpheader{}
\makeatother

\makeatletter
\@addtoreset{equation}{section}
\makeatother

\newbox{\ORCIDicon}
\sbox{\ORCIDicon}{\large \includegraphics[width=0.8em]{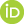}}

\abstract{
Pseudo-Nambu-Goldstone (pNG) dark matter (DM) is a promising DM candidate  
and able to explain the measured DM abundance by the thermal freeze-out mechanism evading the stringent bound from DM direct detection experiments.
We propose a new model providing a pNG DM
by introducing two Standard-Model-singlet complex scalars with the same charges of a dark $U(1)$ gauge symmetry.
They are also charged under a $U(1)$ global symmetry corresponding to their relative phase rotations,
which is explicitly broken by a soft-breaking term in the scalar potential.
The both $U(1)$ symmetries are spontaneously broken by their vacuum expectation values, 
giving rise to one real pNG boson.
We also introduce a discrete $\mathbb{Z}_2$ symmetry exchanging the two scalars to stabilize the pNG boson as DM.
It is shown that this model reproduces the DM abundance consistently with the current bound from the direct detection experiments.
The model has a gauge kinetic mixing between the dark and $U(1)_Y$ gauge fields,
which allows the dark gauge boson to decay even with a relatively light mass
and prevents it from being an additional DM component.
The Landau pole is avoided thanks to the small gauge coupling constant. 
In addition, a DM pair dominantly annihilates into a pair of the dark gauge bosons if the gauge boson mass is lighter than the DM mass,
and thus its cross section 
has significantly different parameter dependence from other pNG DM models.
We also calculate the DM-nucleon scattering cross section at the loop level.
It turns out that it is necessary to probe region covered by the neutrino fog in order to test this model.
}

\begin{document}
\preprint{KEK-TH-2589, DESY-24-002, KYUSHU-HET-277}

\title{A model of pseudo-Nambu–Goldstone dark matter with two complex scalars}

\author[a]{Tomohiro~Abe,\,\href{https://orcid.org/0000-0003-0849-3244}{\usebox{\ORCIDicon}}}
\affiliation[a]{Department of Physics and Astronomy, Faculty of Science and Technology, Tokyo University of Science, Yamazaki, Noda,
Chiba 278-8510, Japan}
\emailAdd{abe.tomohiro(at)rs.tus.ac.jp}

\author[b,c,d]{Yu~Hamada,\,\href{https://orcid.org/0000-0002-0227-5919}{\usebox{\ORCIDicon}}}
\affiliation[b]{Deutsches Elektronen-Synchrotron DESY, Notkestr. 85, 22607 Hamburg, Germany}
\affiliation[c]{KEK Theory Center, 1-1 Oho, Tsukuba, Ibaraki 305-0801, Japan}
\affiliation[d]{Research and Education Center for Natural Sciences, Keio University, 4-1-1 Hiyoshi, Yokohama, Kanagawa 223-8521, Japan}
\emailAdd{yu.hamada(at)desy.de}

\author[e]{and Koji~Tsumura\,\href{https://orcid.org/0000-0003-3765-2750}{\usebox{\ORCIDicon}}}
\affiliation[e]{Department of Physics, Kyushu University,
744 Motooka, Nishi-ku, Fukuoka, 819-0395, Japan}
\emailAdd{tsumura.koji(at)phys.kyushu-u.ac.jp}

\maketitle

\section{Introduction}

It has been a long-standing problem over decades to discover dark matter (DM) in our universe and elucidate its nature. Among many theoretical models for DM, one of the most promising frameworks is the weakly interacting massive particle (WIMP) scenario, in which DM particles interact with the standard model (SM) particles and were in the thermal bath made of them in the early universe. The interaction rate decreases as the universe expands, and the interactions are eventually decoupled, resulting in the DM number density per comoving volume being fixed (frozen out) as the thermal relic abundance. To explain the measured value of the DM energy density~\cite{1807.06209} the annihilation cross section of the DM into the SM particles is about $\langle\sigma v \rangle \simeq 10^{-26}\, \mr{cm}^3 \,\mr{s}^{-1}$ as the thermal averaged value. On the other hand, the crossing symmetry of the Feynman diagrams implies that the elastic scatterings between the DM and SM particles also happen. Such scattering processes have been tried to be  detected by the DM direct detection experiments. Nevertheless, there are no clear signals of the scatterings yet, which imposes stringent upper bounds on the DM-nucleon scattering cross section~\cite{PandaX-4T:2021bab,LUX2022,XENON:2023cxc}. For the WIMP scenario to work, it is necessary to suppress the DM-SM scattering process while keeping the DM annihilation cross section.

The pseudo-Nambu-Goldstone (pNG) DM models can easily explain the null result of the direct detection experiments while keeping the desired annihilation cross section. The original pNG DM model was proposed in Ref.~\cite{Gross:2017dan}, in which an SM-singlet complex scalar is introduced being charged under a softly-broken global $U(1)$ symmetry. The symmetry is spontaneously broken by a vacuum expectation value (VEV) of the new scalar giving rise to a pNG boson, which obtains its mass from the soft-breaking term and can be regarded as DM. The DM-nucleon scattering cross section is highly suppressed in the limit of zero momentum transfer as a consequence of the NG low-energy theorem (soft pion theorem).

Although the suppression mechanism of the original pNG DM model is quite simple and works well, the model suffers from the domain wall (DW) problem and hence requires some extension or needs the low-scale cosmic inflation to dilute the DWs. Besides the spurion-like extension discussed in Ref.~\cite{Gross:2017dan}, other pNG DM models have been also proposed~\cite{2001.03954,2001.05910,Abe:2021byq,Okada:2021qmi,Abe:2022mlc,Liu:2022evb,Otsuka:2022zdy} to overcome this problem. (See also Refs.~\cite{Cai:2021evx,Abe:2021nih,Abe:2021vat,Cho:2023hek,Maji:2023fba}.)
As shown in Table~\ref{tab:table-models}, they can be classified in terms of several points: whether the DW problem arises or not, the DM is stable or decaying, the DM is a real or complex scalar, and coupling constants remain perturbative until very high-energy scale or not (i.e., the Landau pole). If the DM can decay, one must introduce a much higher energy scale than the DM mass in order to make its lifetime longer than the cosmic age. In addition, the appearance of the Landau pole indicates breakdown of the model at a cutoff scale and the necessity of an ultraviolet (UV) completion. If the cutoff scale is quite low, the model is not efficient as an effective theory and the predictive power is limited. Indeed, the model in Ref.~\cite{Abe:2022mlc} has a large dark gauge coupling constant, which soon gets into a non-perturbative regime at a higher energy scale by the evolution of the renormalization group. This requires a UV completion by, e.g., embedding the Abelian gauge group into a non-Abelian gauge group as studied in Ref.~\cite{Otsuka:2022zdy}.

\begin{table}[tbp]
\centering
\begin{tabular}{c|c|c|c|c}
                   & DW problem & stable/decaying & real/complex& Landau pole \\
\hline \hline 
\cite{Gross:2017dan} &  $\times$ & stable  & real & $\checkmark$ \\
\hline
\cite{2001.03954,2001.05910} &  $\checkmark$ & decaying & real &$\checkmark$ \\
\hline 
\cite{Abe:2021byq,Okada:2021qmi} &  $\checkmark$ & decaying & real &$\checkmark$ \\
\hline 
\cite{Liu:2022evb} &  $\checkmark$ & decaying & real &$\checkmark$ \\
\hline 
\cite{Abe:2022mlc} & $\checkmark$ & stable & complex &$\times$\\
\hline 
\cite{Otsuka:2022zdy} &$\checkmark$ & stable & complex & $\checkmark$\\
\hline 
this work & $\checkmark$ & stable & real & $\checkmark$
\end{tabular}
\caption{
Table for pNG DM models.
The rows correspond to the proposed pNG DM models while
the columns correspond to the following items:
whether the DW problem arises ($\times$) or not ($\checkmark$),
DM is stable or decaying,
DM is a real or complex scalar, and
coupling constants hit the Landau pole soon ($\times$) or not ($\checkmark$).
The last row corresponds to this work.
}
\label{tab:table-models}
\end{table}

In this work, we propose a new pNG DM model,
in which we introduce two SM-singlet complex scalars, and they are charged under two $U(1)$ symmetries:
a dark gauge $U(1)$ symmetry corresponding to their simultaneous phase rotation and a global $U(1)$ symmetry which rotates their relative phases. The latter is explicitly broken by a mass-dimension-two term in the scalar potential, where the other soft-breaking terms are forbidden thanks to the gauge $U(1)$ symmetry.
We further introduce a discrete $\mathbb{Z}_2$ symmetry corresponding to the interchange of two scalars.
After the two $U(1)$ symmetries are spontaneously broken by the VEVs of the scalars, a pNG boson arises from the (softly-broken) global $U(1)$ symmetry and obtains its mass from the soft-breaking term. 
We show that the DM-nucleon scattering cross section vanishes at the tree level,
while the model explains the correct value of the DM relic abundance by the freeze-out mechanism.

The imposed exchange symmetry plays a role to make the setup simple
because it stabilizes the pNG DM without any other additional symmetries such as an unbroken $U(1)$ symmetry in Refs.~\cite{Abe:2022mlc,Otsuka:2022zdy},
leading to the real DM instead of complex one. 
Therefore, this model is the first example of a pNG DM model without the DW problem and with the real and stable DM,
as shown in Table~\ref{tab:table-models}.
Furthermore, a non-zero gauge kinetic mixing term between the dark and SM gauge fields does not spoil the model because the DM is already stabilized and the DM-nucleon scattering processes mediated by the gauge bosons are not allowed at the tree level.
This is a bonus of the exchange symmetry.
This non-zero gauge kinetic mixing has two benefits:
Firstly, it makes the dark gauge boson unstable to decay into the SM particles independently of the mass. 
Thus, the dark gauge coupling constant can be small and does not meet the Landau pole until very high energy scale. 
Secondly, it leads to different phenomenology compared to the conventional pNG DM models.
Indeed, the light dark gauge boson makes the annihilation cross section of a DM pair relatively large in heavier DM mass regimes since the annihilation into them is kinematically allowed,
which results in a relatively large VEV in the dark sector
and makes parameter space to explain the DM abundance significantly different from other pNG DM models.

While the DM-nucleon scattering cross section is suppressed at the tree level,
it is induced at the loop level.
We also calculate the loop-level spin-independent cross section in direct detection experiments.
It turns out that the signal will be hidden by the neutrino fog in favorable parameter space.


This paper is organized as follows.
In Sec.~\ref{sec:improved-model}, we introduce our model containing the two complex scalars and the dark gauge boson
with two $U(1)$ symmetries and the exchange symmetry.
We discuss experimental and theoretical constraints on the model in Sec.~\ref{sec:constraints}.
In Sec.~\ref{sec:relic-abundance}, we show that the DM relic abundance of this model as the thermal relic abundance.
In Sec.~\ref{sec:loop}, we discuss the loop effect for the DM direct detection experiments. The details of the loop calculation are given in Appendix~\ref{app:loop}. 
Section~\ref{sec:conclusion} is devoted to the discussion and conclusion.
In Appendix~\ref{sec:appendix}, we discuss a relation between the model with the previous model given in Ref.~\cite{Abe:2022mlc}.
In Appendix~\ref{sec:U1xU1model}, we present a naive version of the pNG DM model without the exchange symmetry and show that it leads to the sizable DM-nucleon scattering cross section in general.

\section{The model}
\label{sec:improved-model}

\subsection{Lagrangian}
We introduce two complex scalars $\phi_1$ and $\phi_2$, 
which are SM singlets and transform under a gauged $U(1)$ rotation called $U(1)_V$,
\begin{equation}
 \phi_1 \to e^{i\alpha(x)} \phi_1, \quad \phi_2 \to e^{i\alpha(x)} \phi_2 \, ,
\end{equation}
where $\alpha(x)$ is an arbitrary real-valued function.
We also impose a softly-broken $U(1)$ global symmetry defined as
\begin{equation}
  \phi_1 \to e^{i\theta } \phi_1, \quad \phi_2 \to e^{-i\theta} \phi_2,
\end{equation}
which is the relative phase rotation and is called the $U(1)_a$ symmetry.
Furthermore, we introduce a discrete $\mathbb{Z}_2$ symmetry under the exchange of the scalars,
\begin{equation}
 \phi_1 \leftrightarrow \phi_2 \,.
\end{equation}
Under this exchange symmetry, the linear combination 
\begin{equation}
 \phi_-\equiv \frac{\phi_1 - \phi_2}{\sqrt{2}} \label{eq:phi-}
\end{equation}
 flips the overall sign, while 
\begin{equation}
  \phi_+\equiv \frac{\phi_1 + \phi_2}{\sqrt{2}} \label{eq:phi+}
\end{equation}
and all other fields do not. This is nothing but the $\mathbb{Z}_2$ symmetry that stabilizes particles arising from $\phi_-$.

The Lagrangian of our model is given as
\begin{equation}
 \mathcal{L}=  \mathcal{L}_\text{\{SM w/o Higgs potential\}} +|D_\mu \phi_1|^2 + |D_\mu \phi_2|^2 - \frac{1}{4}V^{\mu\nu}V_{\mu\nu} - \frac{\sin\epsilon}{2}V^{\mu\nu}Y_{\mu\nu} - V(H,\phi_1 , \phi_2)\label{eq:lagrangian}
\end{equation}
with the scalar potential
\begin{align}
V (H,\phi_1 , \phi_2) & = m_1^2 \left(|\phi_1|^2 + |\phi_2|^2\right)  - \left(m_{12}^2 \phi_1^\ast \phi_2 + \mr{h.c.}\right) \n\label{eq:potential1} \\
&\hspace{3em} + \f{\lambda_1}{2} \left(|\phi_1|^4 + |\phi_2|^4 \right) + \lambda_3 |\phi_1|^2 |\phi_2|^2 \n \\
&\hspace{3em} - m_H^2 |H|^2 + \lambda_H |H|^4 + \lambda_{H1} |H|^2 \left(|\phi_1|^2 + |\phi_2|^2 \right)
\end{align}
and the covariant derivatives
\begin{equation}
 D_\mu \phi_i  = (\partial_\mu - i g_V^{} V_\mu) \phi_i \hspace{2em} (i=1,2),
\end{equation}
where $H$ is the Higgs doublet in the SM. 
Here $V_{\mu\nu}$ and $Y_{\mu\nu}$ are the field strengths of a dark $U(1)_V$ gauge field $V_\mu$ and the SM $U(1)_Y$ gauge field $Y_\mu$.
In general, $V_\mu$ has a gauge kinetic mixing with $Y_\mu$ proportional to the mixing parameter $\sin\epsilon$.
The $U(1)_a$ symmetry is explicitly and softly broken only by the parameter $m_{12}^2$ in the scalar potential.
Other soft-breaking terms such as $\phi_1 \phi_1+\phi_2 \phi_2$ are forbidden by the $U(1)_V$ gauge symmetry.

\subsection{VEVs and stationary condition}
In the following, we assume the scalar fields to take VEVs as follows,
\begin{align}
 \expval{H} = \f{1}{\sqrt{2}}
\begin{pmatrix}
 0 \\ v 
\end{pmatrix}, \quad
\expval{\phi_1} = \f{v_s}{2} ,\quad
\expval{\phi_2} = \f{v_s}{2} ,
\label{eq:VEVs}
\end{align}
which spontaneously break both the symmetries, $U(1)_V$ and $U(1)_a$.
We have assumed that the exchange symmetry (or equivalently, $\phi_\pm \to \pm\phi_\pm$ symmetry) is not broken at the vacuum, making the VEVs of $\phi_1$ and $\phi_2$ to be equal.
Here one can take the VEVs as real without loss of generality by redefining the fields.
The NG boson from the spontaneous $U(1)_V$ symmetry breaking is eaten by the gauge field $V_\mu$.
On the other hand, the spontaneous breaking of $U(1)_a$ produces a NG boson denoted by $a$ that acquires mass due to the explicit breaking term proportional to $m_{12}^2$,
and hence can be used as the pNG dark matter as discussed below.

We obtain the following stationary conditions for the VEVs given in Eq.~\eqref{eq:VEVs}:
\begin{align}
\begin{cases}
m_H^2 = \frac{1}{2} \lambda_{H1} v_s^2 + \lambda_H v ^2 \\
\mathrm{Im} \,m_{12}^2 =0 \\
m_1^2 = \mathrm{Re}\, m_{12}^2  - \f{1}{2}\lambda_{H1} v ^2 - \f{1}{4}\lambda_1 v_s^2 - \f{1}{4}\lambda_3  v_s^2 
\end{cases}
\end{align}
from which it follows that $m_{12}^2$ should be real.

\subsection{Scalar mass spectrum}

There are four real scalar particles around the vacuum as
\begin{equation}
 H = \f{1}{\sqrt{2}}
\begin{pmatrix}
 0 \\ v  + \sigma(x)
\end{pmatrix}\, , 
\end{equation}
\begin{equation}
\phi_1 = \f{1}{\sqrt{2}} \left(\frac{v_s}{\sqrt{2}} + s_1(x) + i \frac{a (x)}{\sqrt{2}} \right), \quad
\phi_2 = \f{1}{\sqrt{2}} \left(\frac{v_s}{\sqrt{2}} + s_2(x) - i \frac{a(x)}{\sqrt{2}}\right) \, ,
\end{equation}
where we have taken the unitary gauge for the $U(1)_V$ gauge and the SM gauge symmetries. The direction of $a$ is orthogonal to $U(1)_V$.
Note that $s_1$ and $s_2$ are not mass eigenstates but mix with each other.
In addition, due to the scalar portal couplings $\lambda_{H1}$, they mix with $\sigma$ from the SM Higgs doublet.

To discuss the $\mathbb{Z}_2$ charges of the particles, 
it is convenient to define new fields as
\begin{equation}
 s_+ \equiv \frac{s_1 + s_2}{\sqrt{2}} , \quad  s_- \equiv \frac{s_1 - s_2}{\sqrt{2}} \, .
\end{equation}
From this, it is obvious that $\phi_+$ and $\phi_-$ defined by Eqs.~\eqref{eq:phi-} and \eqref{eq:phi+} are expressed as 
\begin{equation}
\phi_+= \frac{ v_s + s_+}{\sqrt{2}} \, ,
\end{equation}
\begin{equation}
 \phi_-= \frac{ s_- + ia}{\sqrt{2}} \, ,
\end{equation}
and hence $s_-$ and $a$ are $\mathbb{Z}_2$-odd fields under the exchanging symmetry $\phi_1\leftrightarrow\phi_2$ (or $\phi_\pm \to \pm \phi_\pm$)
while $s_+$ and all the other particles are $\mathbb{Z}_2$-even.

Let us look at the scalar mass spectrum.
Because $a$ is not mixed with the other particles thanks to the CP symmetry in the scalar sector,
the mass of $a$ is easily read off as
\begin{equation}
  \mathcal{L} \supset 
-  \, m_{12}^2 ~a^2 \equiv - \f{1}{2}  m_\mathrm{DM}^2\, a^2 .
\end{equation}
For the other scalars, 
we get the mass matrix 
\begin{align}
 \mathcal{L} \supset 
- \f{1}{2}
\begin{pmatrix}
 \sigma & s_+ & s_-
\end{pmatrix}
{M}^2_\mr{even}
\begin{pmatrix}
 \sigma \\ s_+ \\ s_-
\end{pmatrix}
\end{align}
with
\begin{equation}
 {M}^2_\mr{even} = 
 \begin{pmatrix}
  2\lambda_H v ^2 & \lambda_{H1} v  v_s &  0 \\
   \lambda_{H1} v  v_s & (\lambda_1+\lambda_3) v_s^2/2 & 0 \\
   0 & 0 & 2 m_{12}^2 + (\lambda_1-\lambda_3) v_s^2/2
 \end{pmatrix} \, .
\end{equation}
Note that $s_-$ is not mixed with the other scalars thanks to the exchange symmetry.
The submatrix for $s_+$ and $\sigma$ is further diagonalized in terms of the mass eigenstates
\begin{align}
\begin{pmatrix}
 h \\ h'
\end{pmatrix}
\equiv
 \begin{pmatrix}
  \cos\xi & \sin \xi \\
  -\sin\xi & \cos \xi
 \end{pmatrix}
\begin{pmatrix}
 \sigma \\s_+
\end{pmatrix} \, ,
\end{align}
where $\xi$ is the mixing angle.
Thus we obtain 
\begin{align}
 \mathcal{L} \supset 
- \f{1}{2} m_h^2 \,h^2
 - \f{1}{2} m_{h'}^2 \,h'^2 
 - \f{1}{2} m_{s_-}^2 \, s_-^2 
\end{align}
with the mass eigenvalues
\begin{align}
 \begin{pmatrix}
  m_h^2 & 0 \\
  0 & m_{h'}^2
 \end{pmatrix}
&\equiv
  \begin{pmatrix}
  \cos\xi & \sin \xi \\
  -\sin\xi & \cos \xi
 \end{pmatrix}
 \begin{pmatrix}
  2\lambda_H v ^2 & \lambda_{H1} v  v_s  \\
   \lambda_{H1} v  v_s & (\lambda_1+\lambda_3) v_s^2/2 
 \end{pmatrix} 
  \begin{pmatrix}
  \cos\xi & -\sin \xi \\
  \sin\xi & \cos \xi
 \end{pmatrix}
,\\ \nonumber\\
m_{s_-}^2 &\equiv 2 m_{12}^2 + (\lambda_1-\lambda_3) v_s^2/2 \, .
\end{align}
We take the mass eigenstate $h$ as the SM-like Higgs boson,  and thus $m_h=125$~GeV.

Collecting the stationary conditions with respect to the VEVs and the definitions of the mixing angle given above,
we can express the parameters in the potential in terms of the mass eigenvalues, mixing angle, and VEVs as
\begin{align}
 \lambda_H & = \frac{ m_{h}^2 \cos \xi^2 +   m_{h'}^2 \sin \xi^2}{2 v^2},\label{eq:parameter-pot-1} \\
 \lambda_{H1} & = \frac{(m_{h}^2 - m_{h'}^2) \cos \xi \sin \xi}{v v_s}, \\
 \lambda_1 & =  \frac{ m_{h}^2 \sin \xi^2 + m_{h'}^2  \cos \xi^2}{v_s^2} + \frac{m_{s_-}^2-m_\mr{DM}^2}{v_s^2}, \\
 \lambda_3 & =  \frac{ m_{h}^2 \sin \xi^2 + m_{h'}^2  \cos \xi^2}{v_s^2} - \frac{m_{s_-}^2-m_\mr{DM}^2}{v_s^2},\label{eq:parameter-pot-2} \\
 m_1^2 &= \frac{1}{2} m_\mr{DM}^2 - \f{1}{2}\lambda_{H1} v^2 - \f{1}{4}\lambda_1 v_s^2 - \f{1}{4}\lambda_3  v_s^2, \\
 m_H^2 & = \frac{1}{2} \lambda_{H1} v_s^2 + \lambda_H v^2,  \\
\mathrm{Im} \,m_{12}^2 & =0, \\
\mathrm{Re} \,m_{12}^2 & = \frac{1}{2} m_\mr{DM}^2 \, .
\end{align}
From this, one can see that $s_-$ and $a$ have degenerated masses if and only if $\lambda_1=\lambda_3$,
where the exchange symmetry and the two $U(1)$ symmetries are merged and extended into a (softly-broken) global $O(4)$ symmetry acting on $(\mr{Re}\, \phi_1, \mr{Im}\, \phi_1,\mr{Re}\, \phi_2,\mr{Im}\, \phi_2)$,
resulting in that the model becomes equivalent to that in Ref.~\cite{Abe:2022mlc}.
See Appendix.~\ref{sec:appendix} for more detail.

\subsection{Stability of DM and gauge kinetic mixing}
Since $a$ and $s_-$ have the odd charges under the exchange symmetry,
the lighter one can be a stable particle without introducing any other symmetries.
Therefore, as far as taking $m_{s_-}>m_\mr{DM}^{}$, we have the stable pNG DM $a$.
Note that taking $m_{s_-}^{}<m_\mr{DM}^{}$ makes $s_-$ the main DM component, 
and hence the model effectively reduces to a model with one real DM $s_-$ and two real scalar mediators ($h$ and $h'$ defined above).
Such a model is easily excluded by the direct detection experiments in most parameter space~\cite{Arcadi:2017kky,Arcadi:2019lka}.

In contrast to the model in Ref.~\cite{Abe:2022mlc},
the dark CP symmetry is not required to stabilize the pNG DM,
which allows the model to have the non-zero gauge kinetic mixing, $\sin \epsilon \neq 0$.
It mixes the SM $U(1)_Y$ gauge field $Y_\mu$ and the dark gauge field $V_\mu$.
It is convenient to introduce a new basis
\begin{equation}
 \begin{pmatrix}
  Y'_\mu \\ V'_\mu
 \end{pmatrix}
\equiv
 \begin{pmatrix}
  1 & \sin\epsilon \\ 0 &\cos\epsilon
 \end{pmatrix}
 \begin{pmatrix}
  Y_\mu \\ V_\mu
 \end{pmatrix} \, ,
\end{equation}
which modifies the gauge kinetic terms into canonical forms, 
\begin{equation}
 - \frac{1}{4}V^{\mu\nu}V_{\mu\nu} - \frac{\sin\epsilon}{2}V^{\mu\nu}Y_{\mu\nu} - \frac{1}{4}Y^{\mu\nu}Y_{\mu\nu}
=
 - \frac{1}{4}V'^{\mu\nu}V'_{\mu\nu} -\frac{1}{4}Y'^{\mu\nu}Y'_{\mu\nu} \, .
\end{equation}
In this new basis, the covariant derivatives for the scalars become
\begin{align}
 D_\mu H &= \left(\partial_\mu - i \frac{g}{2} W_\mu^a \sigma^a  - i \frac{g_Y^{}}{2} Y'_\mu + i \tan\epsilon\frac{g_Y^{}}{2} V'_\mu \right) H, \\
 D_\mu \phi_i  &= \left(\partial_\mu - i \frac{1}{\cos\epsilon} g_V^{} V'_\mu\right) \phi_i, \hspace{2em} (i=1,2).
\end{align}
In the vacuum, the VEVs of the scalar fields give masses to the gauge bosons as
\begin{align}
 \mathcal{L}&
\supset 
\frac{g_Z^2 v^2}{8}
\begin{pmatrix}
 W_\mu^3 & Y_\mu' & V_\mu'
\end{pmatrix} 
\begin{pmatrix}
 c_w^2 & - c_w s_w & c_w s_w t_\epsilon \\
 -c_w s_w & s_w^2 & -s_w^2 t_\epsilon \\
 c_w s_w t_\epsilon & -s_w^2 t_\epsilon & t_\epsilon^2 s_w^2 + R^2/c_\epsilon^2
\end{pmatrix}
\begin{pmatrix}
 W_\mu^3 \\ Y_\mu' \\ V_\mu'
\end{pmatrix}, \label{eq:mass-matrix}
\end{align}
where
$R \equiv 2g_V^{} v_s /(g_Z^{} v)$, $g_Z^{} \equiv\sqrt{g^2 + g_Y^2}$, $s_w=\sin \theta_W^{}= g_Y^{}/g_Z^{}$, $c_w=\cos \theta_W^{}= g/g_Z^{}$,
$c_\epsilon\equiv\cos \epsilon$, and $t_\epsilon \equiv \tan \epsilon$.
We introduce the mass eigenbasis $(Z_\mu, A_\mu,Z'_\mu)$ as
\begin{align}
 \begin{pmatrix}
 W_\mu^3 \\ Y_\mu' \\ V_\mu'
\end{pmatrix}
&= 
\begin{pmatrix}
 c_\zeta & 0 & s_\zeta \\
 0 & 1 & 0 \\
 -s_\zeta & 0 & c_\zeta
\end{pmatrix}
\begin{pmatrix}
 c_w & s_w & 0 \\
 -s_w & c_w & 0 \\
 0 & 0 & 1
\end{pmatrix}
\begin{pmatrix}
 Z_\mu \\ A_\mu \\ Z'_\mu
\end{pmatrix} 
=
\begin{pmatrix}
 c_\zeta c_w& c_\zeta s_w & s_\zeta \\
 -s_w & c_w & 0 \\
 -s_\zeta c_w & -s_\zeta s_w & c_\zeta
\end{pmatrix}
\begin{pmatrix}
 Z_\mu \\ A_\mu \\ Z'_\mu
\end{pmatrix}  
,
\end{align}
where $\zeta$ satisfies
\begin{equation}
 \tan 2\zeta = \frac{2 s_\epsilon c_\epsilon s_w}{R^2 -1 + s_\epsilon^2(1+s_w^2)}.
\end{equation}
In this basis, the mass matrix in Eq.~\eqref{eq:mass-matrix} is diagonalized, and Eq.~\eqref{eq:mass-matrix} is given by
\begin{align}
\frac{1}{2} m_Z^2 \, Z_\mu Z^\mu 
+\frac{1}{2} m_{Z'}^2 \, Z'_\mu Z'^\mu \, ,
\end{align}
where
\begin{align}
 m_Z^2 &= \frac{g_Z^2 v^2}{4} \left(1- t_\epsilon s_w \sin 2\zeta + s_\zeta^2 \left(\frac{R^2}{c_\epsilon^2} + t_\epsilon^2 s_w^2 -1\right)\right), \\
m_{Z'}^2 &= \frac{g_Z^2 v^2}{4} \left(1 + t_\epsilon s_w \sin 2\zeta + c_\zeta^2 \left(\frac{R^2}{c_\epsilon^2} + t_\epsilon^2 s_w^2 -1\right)\right) \, .\label{eq:Z'mass}
\end{align}

Now we can express nine parameters in the scalar and the new gauge sectors
\begin{align}
 \lambda_H, ~\lambda_1, ~\lambda_3, ~\lambda_{H1}, ~ m_1, ~ m_H^{} ,~ \mr{Re}\, m_{12}^2, ~ g_V^{}, ~ \sin \epsilon \, ,
\end{align}
 in terms of the following physical parameters
\begin{align}
v(=246 \, \mr{GeV}),\ m_{h} (=125 \, \mr{GeV}), ~
 v_s, ~ m_{h'} ,~ m_{s_-}, ~ m_\mr{DM}^{} , ~ m_{Z'}^{}, ~ \sin \xi, ~\sin\epsilon \, .
\end{align}

\subsection{Suppression of scattering cross section in direct detection}\label{sec:DDatTree}


We here confirm that the spin-independent cross section of the pNG DM $a$ with nucleons is indeed suppressed in our model.\footnote{
In contrast to the other pNG DM models, 
the fact that the soft-breaking term has the mass dimension two does not ensure the vanishment of the scattering cross section.
Instead, the exchange symmetry of $\phi_1$ and $\phi_2$ also plays a crucial role in the vanishment.
One can check that it indeed does not vanish in general models without the exchange symmetry.
To suppress the cross section, one needs a fine-tuning for the VEVs of the scalars $\phi_1$ and $\phi_2$ to be equal.
See Appendix~\ref{sec:U1xU1model} for the details.
}
To see this, it is convenient to move on to the non-linear representation, 
\begin{align}
 \phi_1 &= \frac{1}{\sqrt{2}} \left(\frac{v_s}{\sqrt{2}} +s_1\right) \exp\left[i \frac{\pi_a}{v_s}\right] \, ,\\
 \phi_2 &= \frac{1}{\sqrt{2}} \left(\frac{v_s}{\sqrt{2}} +s_2\right) \exp\left[-i \frac{\pi_a}{v_s}\right]  \, ,
\end{align}
in which $\pi_a$ corresponds to $a$ in the linear representation and we have taken the unitary gauge again.
In this representation, the vanishment of the cross section is caused by the vanishment of the cubic couplings $\pi_a \pi_a s_1$ and $\pi_a \pi_a s_2$ instead of non-trivial cancellation between different diagrams, as is studied in Ref.~\cite{Cai:2021evx}.

To read off the cubic couplings, 
we substitute these expressions into the Lagrangian and obtain
%
%
%
\begin{align}
 & |\partial_\mu \phi_1|^2 +  |\partial_\mu \phi_2|^2 + m_{12}^2 \left(\phi_1^\ast \phi_2 + \mathrm{h.c.}  \right)\\
& \supset 
\frac{1}{2 } \left[\left(\frac{v_s}{\sqrt{2}} +s_1\right)^2 + \left(\frac{v_s}{\sqrt{2}} +s_2\right)^2 \right]\left(\frac{\partial_\mu \pi_a}{v_s}\right)^2
 \nonumber \\
& \hspace{3em}-  
\frac{1}{2} 2 m_\mathrm{DM}^2 \left(\frac{v_s}{\sqrt{2}} + s_1\right)\left(\frac{v_s}{\sqrt{2}} + s_2\right)\left(\frac{\pi_a}{v_s}\right)^2 \, ,
\end{align}
from which we get the cubic couplings,
\begin{align}
\mathcal{L} &\supset
-\frac{\sqrt{2} }{v_s} \left[
\left(s_1+s_2\right)\pi_a \left(\partial_\mu\partial^\mu+m_\mr{DM}^2\right) \pi_a
+ \pi_a \partial^\mu \pi_a \partial_\mu \left(s_1 + s_2\right)
 \right]\,. \label{eq:aas} 
\end{align}
In the direct detection experiments for the DM-nucleon scattering,
the first term in Eq~\eqref{eq:aas} vanishes due to the on-shell condition for $\pi_a$, $(\partial_\mu \partial^\mu + m_\mathrm{DM}^2) \pi_a=0$,
which results in that the cubic couplings are proportional to the momentum of $s_1$ or $s_2$.
Therefore, in the limit of the zero momentum transfer, the DM-nucleon scattering cross section vanishes at the tree level and is highly suppressed,
which is consistent with the current null results.

Note that it is not difficult to show the suppression of the scattering cross section in the \textit{linear} representation,
where one needs to calculate two diagrams mediated by two mass eigenstates $h$ and $h'$.
The diagrams cancel with each other, which is equivalent to the result shown above.

\section{Constraints}
\label{sec:constraints}

In this section, we discuss constraints from the perturbative unitarity, the Higgs invisible decay, and the scalar mixing angle.

\subsection{Perturbative unitarity}

One may obtain the constraints on the scalar and gauge couplings from the perturbative unitarity (PU) bound~\cite{Lee:1977eg}.
Let us denote the matrix element in two-to-two scattering $i\to f$ by
\begin{equation}
 (2\pi)^4 \delta^{(4)} (P_f-P_i) \mathcal{M}_{fi}(\sqrt{s},\cos\theta) = \bra{f}T\ket{i},
\end{equation}
where $T$ is the interaction part of the $S$-matrix, $S=1 + i T$, and $\theta$ is the scattering angle.
At high-energy scattering $s\to \infty$, $\mathcal{M}_{fi}(\sqrt{s},\cos\theta)$ is expanded by partial waves as
\begin{align}
 {\cal M}_{fi} = 16 \pi \sum_\ell (2 \ell + 1) P_\ell(\cos\theta) a^\ell_{fi} \, ,
\end{align}
where $P_\ell$ is the Legendre polynomial.
Since $\ell=0$ scattering typically gives the most stringent bound, we hereafter focus on $\ell=0$.
From the unitarity of the $S$-matrix $S S^\dagger=1$, 
we have $ i (T - T^\dagger) + T T^\dagger = 0$,
and hence
\begin{equation}
 \frac{i}{2} \left(a_{fi}^0 - (a_{if}^0)^\ast\right) + \sum_j (a_{jf}^0)^\ast \, a_{ji}^0 \leq 0
\end{equation}
with $j$ being all possible two-particle states.
Here the inequality comes from the fact that the left-hand side is underestimated by concentrating on the two-particle states as intermediate states.
The above inequality immediately gives
\begin{equation}
  - \mr{Im}\, \lambda^0 + |\lambda^0|^2 \leq 0,\label{eq:unitarity-circle}
\end{equation}
for each eigenvalue $\lambda^0$ of the matrix $a^0$.
Equation~\eqref{eq:unitarity-circle} means that each eigenvalue should be inside a circle of a radius $1/2$ whose center is at $(\mr{Re}\, \lambda^0 , \mr{Im}\, \lambda^0)=(0,1/2)$.

The above argument implies the perturbative unitarity that all eigenvalues should satisfy at the tree-level the following inequality:
\begin{equation}
 |\mr{Re}\, \lambda^0| \leq \frac{1}{2} \quad \text{for all eigenvalues $\lambda^0$ of $a^0$}.
\end{equation}
We utilize this inequality to find the constraints on the scalar and gauge couplings.

\subsubsection{scalar quartic couplings}
We consider the scalar two-to-two scattering processes $i\to f$ for $\ell=0$.
One can carry out the calculation in the symmetric phase because of the high-energy scattering.
For scatterings between charge-neutral states, i.e.,
$i\to f$ with
$i,f \in \{H_1 H_1^\dagger, \, H_2 H_2^\dagger,\phi_1 \phi_1^\dagger, \, \phi_2 \phi_2^\dagger, \, \phi_1 \phi_2^\dagger \}$,
where $H_1,H_2$ are the upper and lower components of the SM Higgs doublet $H$, 
the matrix $a_0$ is expressed as
\begin{align}
(a^0)_{fi} =   \frac{1}{16\pi} 
\begin{pmatrix}
4\lambda_H & 2 \lambda_H & \lambda_{H1} & \lambda_{H1} & 0\\
2\lambda_H & 4 \lambda_H & \lambda_{H1} & \lambda_{H1} & 0\\
\lambda_{H1} & \lambda_{H1} & 2\lambda_1 & \lambda_3 & 0\\
\lambda_{H1} & \lambda_{H1} & \lambda_3 & 2\lambda_1 & 0\\
0 & 0 & 0 & 0 & \lambda_3
\end{pmatrix}  \, ,
\end{align}
from which we get the PU bound on the eigenvalues,
\begin{align}
 |\lambda_H| <&\, 4\pi,  \label{eq:PU-scalar1} \\
 |2\lambda_1-\lambda_{3}| <& \, 8\pi,\\
 |\lambda_{3}| <& \, 8\pi,\\
 \left|2\lambda_1+\lambda_3 + 6 \lambda_H \pm \sqrt{(2\lambda_1 + \lambda_3 - 6 \lambda_H)^2 +16 \lambda_{H1}^2} \right| <&\, 16\pi \, .
\end{align}
In addition, we have also scattering modes between $U(1)_V$ charged states such as $\phi_1 \phi_1 \to \phi_1 \phi_1$ and $\phi_1 H_1 \to \phi_1 H_1$,
which give the PU bound
\begin{align}
 |\lambda_1|< &\, 8\pi, \\
 |\lambda_{H1}| < & \, 8\pi \, .\label{eq:PU-scalar2}
\end{align}
We impose all the inequalities Eqs.~\eqref{eq:PU-scalar1}-\eqref{eq:PU-scalar2} in our analysis.

\subsubsection{gauge coupling}
We study $V \phi_1\to V \phi_1$ scattering at high energy in the symmetric phase in order to find the constraint on $g_V^{}$. 
We find that the most stringent bound comes from $\ell=0$ scattering, given as
\begin{align}
 a_0^{++} =  a_0^{--} = \frac{2g_V^2}{16\pi} \, ,
\end{align}
where $++$ and $--$ mean the helicities of the initial and final $V$ states.
From this, we obtain the PU bound for the dark gauge coupling constant as
\begin{equation}
    g_V^{} < \sqrt{4\pi} \, .\label{eq:PU-gauge}
\end{equation}

\subsection{Higgs invisible decay}
If $m_\mr{DM}^{} < m_h/2$, the SM-like Higgs boson $h$ can decay into a pair of the DM $aa$.
The decay width is given by
\begin{align}
 \Gamma(h \to aa)
=&
 \frac{1}{32 \pi} \frac{g_{aa h}^2}{m_h} \sqrt{1 - \frac{4 m_\mr{DM}^2}{m_{h}^2}}\,
 \theta(m_{h} - 2 m_\mr{DM}^{})
\end{align}
with 
\begin{equation}
    g_{aah} = v \lambda_{H1} \cos\xi + \frac{1}{2} v_s (\lambda_1 + \lambda_3) \sin \xi \, .
\end{equation}
This process is the Higgs invisible decay 
and is being searched by the ATLAS and CMS experiments. 
Currently, the ATLAS and CMS experiments obtain the upper bound on the branching ratio as 
\begin{align}
 \text{BR}_\text{inv} <  
\begin{cases}
 0.107 & \text{(ATLAS \cite{ATLAS:2023tkt})} \\
 0.15 & \text{(CMS \cite{CMS:2023sdw})}
\end{cases}
\end{align}
at 95\% CL.

\subsection{Constraints on scalar mixing angle} 

One may get constraints on the scalar mixing angle $\xi$ 
as it decreases couplings of the SM-like Higgs boson to the other SM particles with the factor $\cos\xi$.
From the latest study on the Higgs boson couplings~\cite{ATLAS:2022vkf},
the most stringent lower bound is the vector boson coupling, $\kappa_V^{}\gtrsim 0.97$ (95\% CL), 
from which we can read off the bound on $\xi$ as $\sin\xi\lesssim 0.24$.

\subsection{Landau pole}

Taking into account quantum effects, the running coupling constants depend on the renormalization scale and evolve from 
the infrared (IR) to the UV scales as described by renormalization group (RG) equations.
In particular, the gauge and scalar quartic couplings in the dark sector can grow up and might diverge at the UV scale by hitting the Landau pole when they are sufficiently large at the IR scale.
In this subsection, we write down the RG equations for them.
Since $\lambda_{H1}$ is typically quite small compared to the other coupling constants, we ignore it in the RG analysis.
Note that the non-zero gauge kinetic mixing $\sin\epsilon$ is also negligible.
In the $\overline{\mr{MS}}$ scheme, the one-loop RG equations are given as
\begin{align}
(4 \pi)^2 \beta_{\lambda_1} & \equiv (4 \pi)^2 \, \frac{\dd \lambda_1(\mu)}{\dd \ln \mu}= 10 \lambda_1^2 + 2 \lambda_3^2 + 12  g_V^4 - 12 g_V^2 \lambda_1, \\
(4 \pi)^2 \beta_{\lambda_3} & \equiv (4 \pi)^2 \, \frac{\dd \lambda_2(\mu)}{\dd \ln\mu}= 4\lambda_3^2 + 8\lambda_1 \lambda_3 + 12  g_V^4 - 12 g_V^2 \lambda_3, \\
(4\pi)^2 \beta_{g_V^{}} & \equiv (4 \pi)^2  \, \frac{\dd g_V^{}(\mu)}{\dd \ln\mu} = \frac{2}{3} g_V^3
\end{align}
with $\mu$ being the renormalization scale.
One can see that, when $\lambda_1=\lambda_3$, $\beta_{\lambda_1}=\beta_{\lambda_3}$ holds and hence the RG flow does not depart from the $O(4)$ symmetric critical surface in the three dimensional parameter space.

We investigate the energy scale $\mu= \Lambda$ at which the coupling constants become non-perturbative,
that is, any of the inequalities Eqs.~\eqref{eq:PU-scalar1}-\eqref{eq:PU-scalar2} is violated.
We adopt this scale $\Lambda$ as a practical criterion for the energy scale hitting the Landau pole.

\subsection{Boundedness of potential}
\label{sec:boundedness}
We here provide necessary and sufficient conditions for the scalar potential to be bounded from below in an arbitrary direction with large field values.
Since the quadratic terms are irrelevant for the large field values,
we concentrate on the quartic coupling constants
$\lambda_1,~\lambda_3,~\lambda_H,~\lambda_{H1}$ in Eq.~\eqref{eq:potential1}.
We obtain the conditions as follows:
\begin{align}
& \lambda_1>0,~ \lambda_H>0,~ \lambda_1 + \lambda_3 >0, \nonumber\label{eq:boundedness} \\[1ex]
& \lambda_{H1} + \sqrt{2 \lambda_1 \lambda_H}>0 , ~
\lambda_{H1} + \sqrt{(\lambda_1+\lambda_3) \lambda_H}>0 \, .
\end{align}
The derivation is given in Appendix.~\ref{sec:deriv-boundedness}.
We impose these conditions through the whole parameter space investigated in the following sections.

\subsection{Vacuum stability}
The vacuum that we chose in Eq.~\eqref{eq:VEVs} is classically stable since the square of masses of the all particles around the vacuum are positive.
On the other hand, the vacuum cannot be stable quantum mechanically if there is another minimum with lower potential energy.
We also check that such an additional minimum does not appear at the tree level in the following analysis.

\section{Relic abundance}
\label{sec:relic-abundance}
We here  discuss the DM relic abundance.
This model easily explains the correct relic abundance by the thermal freeze-out mechanism,
in which annihilation processes of $aa$ into the SM particles and their inverse ones are equilibrated in the high-temparature universe while they are eventually decoupled by the cosmic expansion as the universe cools down.
Then the abundance of $a$ is freezed out, which explains the DM relic abundance of the present universe.

The pair annihilation of $a$ is mainly given by $s$-channel processes mediated by $h$ and $h'$ into SM particles.
The pair of DM particles also annihilates into pairs of $h h'$, $h'h'$, and $Z'Z'$ if kinematically allowed.
The annihilation into $s_- s_-$ is not effective
because ${s_-}$ is taken to be heavier than $a$ in order not to be the main DM component as stated above.

Note that, a production of $ZZ$ is allowed for $m_\mr{DM}^{}> m_Z^{}$ through the gauge kinetic mixing.
Nevertheless, it cannot be significant because the cross section is proportional to the fourth-power of the kinetic mixing parameter $\sin \epsilon$ and is highly suppressed in most parameter space of interest.

As stated in Sec.~\ref{sec:improved-model}, 
the dark gauge boson $Z'$ cannot be stable but easily decays into SM particles
thanks to the non-zero value of $\sin \epsilon$, 
which allows us to take a relatively light mass of $Z'$, say, $m_{Z'}^{}=\mathcal{O}(100)$ GeV.
Therefore the annihilation channel into $Z'Z'$ can be easily open,
which results in a relatively large annihilation cross section of $aa$ for $m_\mr{DM}^{}> m_{Z'}^{}$.
This is a big difference compared with a previous model proposed in Ref.~\cite{Abe:2022mlc},
in which the mass of the dark gauge boson must be at least twice as heavy as that of the pNG DM to prevent $Z'$ from becoming the main DM component.

\begin{figure}[tbp]
\centering
\includegraphics[width=0.75\textwidth]{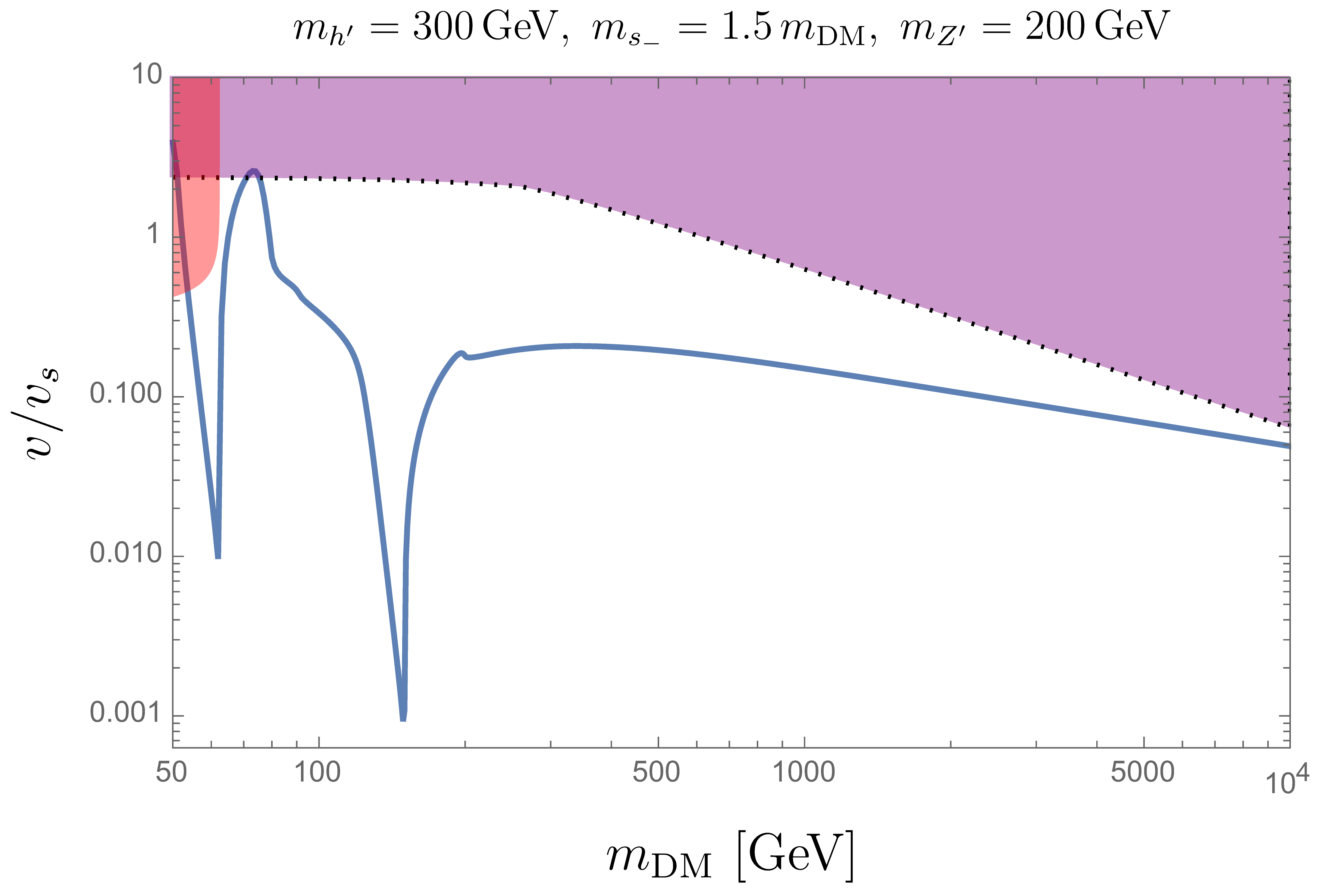}
\caption{
Parameter space to explain the measured DM relic abundance. 
We take $\sin \xi=0.1$, $\sin \epsilon=10^{-4}$, $m_{h'}= 300$ GeV, $m_{s_-}=1.5\, m_\mr{DM}^{}$, and $m_{Z'}^{}=200$ GeV.
The blue thick line indicates the correct relic abundance $\Omega_\mathrm{DM}h^2 = 0.12$.
The red region is excluded by the current bound from the Higgs invisible decay
while the purple region bounded by dotted line is excluded by the PU bound for the scalar quartic and dark gauge coupling, see Sec.~\ref{sec:constraints}.
}
\label{fig:relic}
\end{figure}

As benchmark values, we take the parameters as $\sin \xi=0.1$, $\sin \epsilon=10^{-4}$, $m_{h'}= 300$~GeV, $m_{s_-}=1.5\, m_\mr{DM}^{}$, and $m_{Z'}^{}=200$~GeV.
We use \texttt{micrOMEGAs}~\cite{Belanger:2018ccd} to calculate the DM relic abundance.
Figure~\ref{fig:relic} shows the parameter region giving the correct relic abundance in $m_\mr{DM}^{}$-$(v/v_s)$ plane.
The blue thick line indicates the correct relic abundance $\Omega_\mathrm{DM}h^2 = 0.12$~\cite{1807.06209}.
The red region is excluded by the current bound from the Higgs invisible decay
while the purple region bounded by a dotted line is excluded by the PU bound for the scalar quartic and dark gauge coupling, see Sec.~\ref{sec:constraints}.
We can see that there are two deep spikes for the solid line,
which correspond to the $s$-channel resonances of $h$ and $h'$ at $m_\mr{DM}^{}= 125/2$ GeV and $300/2$ GeV, respectively.
Furthermore, the bound from the Higgs invisible decay is significant for light mass region 
while the PU bound seems insignificant in this model up to $m_\mr{DM}^{}=10$ TeV because $Z'$ is not so heavy that the gauge coupling $g_V^{}$ saturates the PU bound.
We checked that the result is almost independent of the value of $\sin \epsilon$ as long as $\sin \epsilon \lesssim \mathcal{O}(0.1)$ 
because $\sin \epsilon$ appears as $\sin^4 \epsilon$ in the annihilation cross section as stated above.

\begin{figure}[tbp]
\centering
\includegraphics[width=0.75\textwidth]{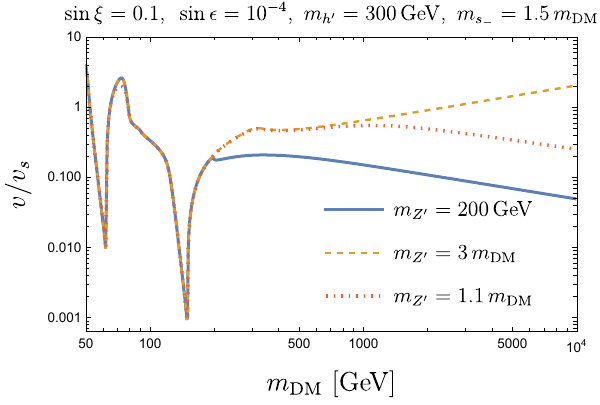}
\caption{
Comparison of three cases of the masses of $Z'$ gauge boson.
The blue thick, orange dashed, and red dotted lines indicate the parameter space providing the measured DM relic abundance
in the cases of $m_{Z'}^{}=200$ GeV, $m_{Z'}^{}=3\,m_\mr{DM}$, and $m_{Z'}^{}=1.1\,m_\mr{DM}^{}$, respectively. 
The other parameters are the same as those in Fig.~\ref{fig:relic}.
The blue thick and the orange dashed ones have distinct asymptotic behavior for heavier $m_\mr{DM}$,
which corresponds to whether the annihilation channel $a a \to Z' Z'$ is open or not.
In the case of $m_{Z'}^{}= 1.1\,m_\mr{DM}^{}$, this channel is partially allowed, resulting in the intermediate behavior of the two lines.
}
\label{fig:comparison}
\end{figure}

Figure~\ref{fig:comparison} shows a comparison of the result of Fig.~\ref{fig:relic} (thick blue curve) to the parameter space in the case that $m_{Z'}^{}$ is fixed as $3\, m_\mr{DM}^{}$ as in Ref.~\cite{Abe:2022mlc} (dashed orange curve)
and to a nearly degenerate case with $m_{Z'}^{}=1.1\, m_\mr{DM}^{}$ (dotted red curve).
The blue one significantly deviates from the orange one for $m_\mr{DM}^{}> m_{Z'}^{}= 200$~GeV, 
in which the annihilation of $aa$ into $Z'Z'$ is a dominant process and hence gives the large cross section.
For the red one,
although $Z'$ is slightly heavier than the DM $a$,
this annihilation channel is still open around the freeze-out temperature
because the Boltzmann factor, $\exp (-(m_{Z}^{} - m_\mr{DM}^{})/T_f )$ ($T_f \simeq m_\mr{DM}^{}/25$),
does not suppress the process for $m_{Z'}^{} \sim m_\mr{DM}^{}$.
Namely, this forbidden channel~\cite{Griest:1990kh} is active for $m_{Z'} = 1.1 m_\mr{DM}$. 
Consequently, the parameter space becomes intermediate of the two curves.
We checked that cases of $m_{Z'}^{}\gtrsim 1.2\, m_\mr{DM}^{}$ are almost the same as that of $m_{Z'}^{}=3\, m_\mr{DM}^{}$ and the prediction becomes independent of $m_{Z'}^{}$.

\begin{figure}[tbp]
\centering
\includegraphics[width=0.75\textwidth]{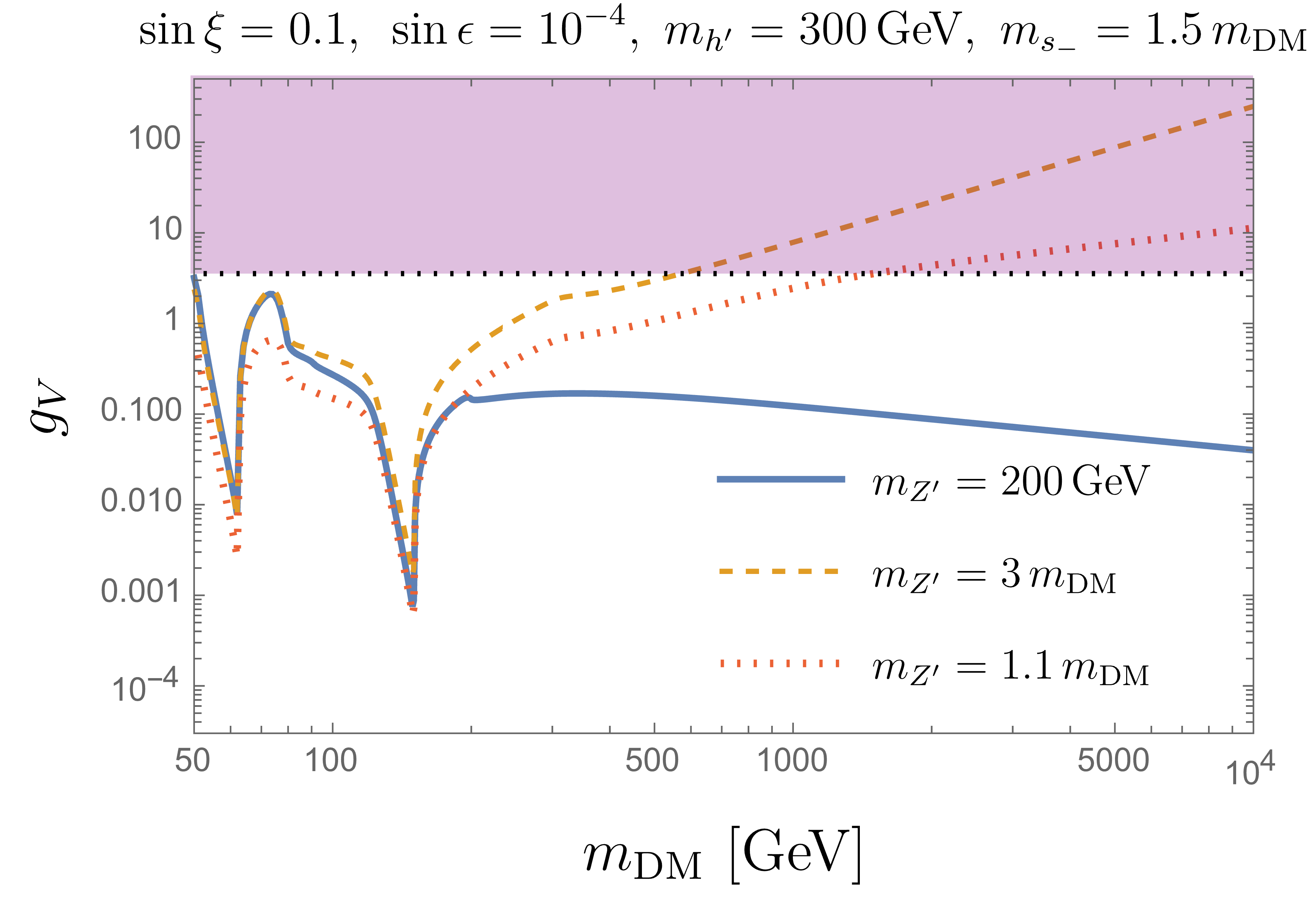}
\caption{
The value of the gauge coupling constant $g_V$ for the parameter space shown in Fig.~\ref{fig:comparison}.
The purple region bounded by dotted line is excluded by the PU bound only for the gauge coupling constant $g_V > \sqrt{4\pi}$.
}
\label{fig:gV}
\end{figure}

In Fig.~\ref{fig:gV}, we plot the value of the gauge coupling constant $g_V^{}$ versus $m_\mr{DM}^{}$ in the both cases for the parameter space shown in Fig.~\ref{fig:comparison}.
The purple region bounded by the dotted line corresponds the PU bound only for the gauge coupling, $g_V^{} > \sqrt{4\pi}$.
One can see that the PU bound on $g_V^{}$ is violated in heavier $m_\mr{DM}^{}$ region in the cases of $m_{Z'}^{}= 3 \, m_\mr{DM}^{} $ and $ 1.1 \, m_\mr{DM}^{}$
while it is not in the case of $m_{Z'}^{}= 200$~GeV.

\begin{figure}[tbp]
\centering
\includegraphics[width=0.65\textwidth]{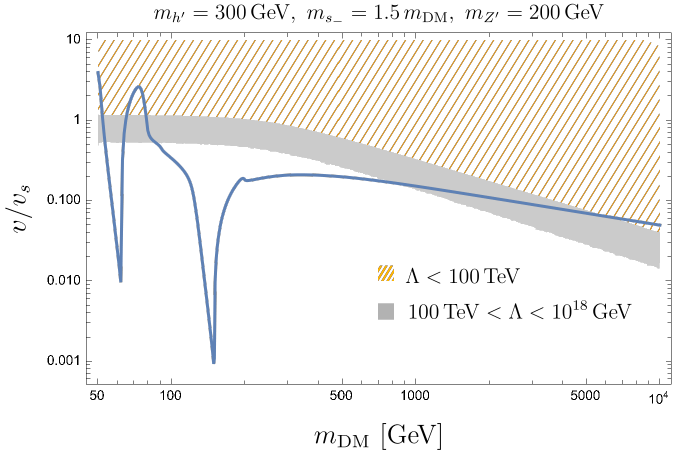} \\[3ex]
\includegraphics[width=0.65\textwidth]{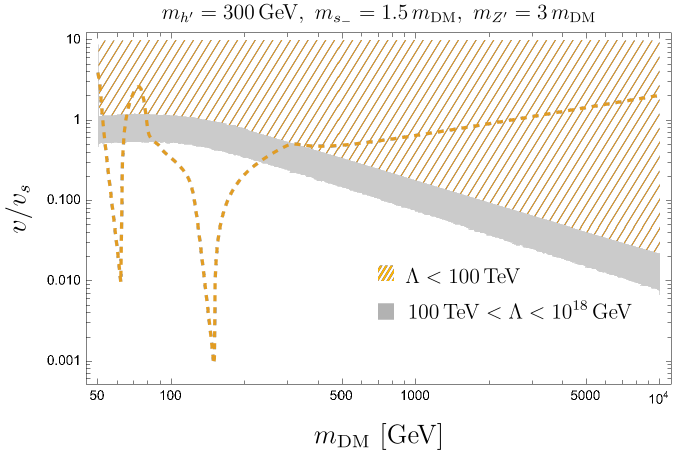}\\[3ex]
\includegraphics[width=0.65\textwidth]{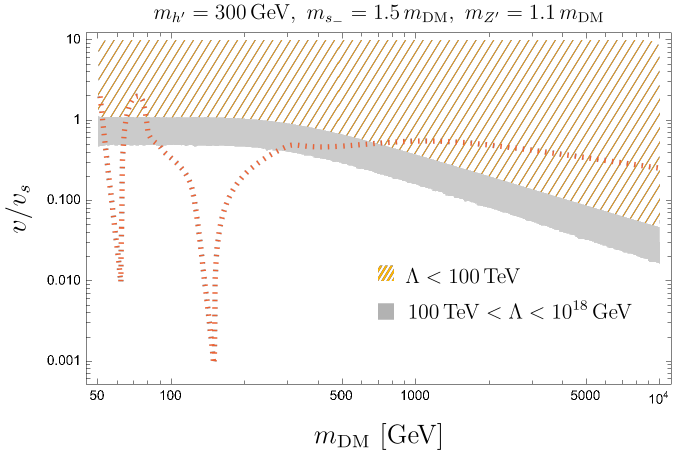}
\caption{
Parameter region in which the running coupling constants grow up into non-perturbative regime (Landau pole) 
at the energy scale $\Lambda<100$ TeV (orange hatched), $100\, \mr{TeV}<\Lambda<10^{18} \, \mr{GeV}$ (gray),
and $\Lambda>10^{18}$ GeV (white).
$m_{Z'}^{}$ is taken as $m_{Z'}^{}=200\, \mr{GeV}$ (top),  $m_{Z'}^{}=3 \, m_\mr{DM}^{}$ (middle), and $m_{Z'}^{}=1.1 \, m_\mr{DM}^{}$ (bottom).
The other parameters are the same as those in Fig.~\ref{fig:relic}.
The curves correspond to those in Fig.~\ref{fig:comparison}.
}
\label{fig:Landau-pole}
\end{figure}

As the results of the RG analysis given in Sec.~\ref{sec:constraints},
Fig.~\ref{fig:Landau-pole} shows parameter region in which any of the running scalar quartic and gauge coupling constants grow up into non-perturbative regime (hitting Landau pole) 
at the energy scale $\mu=\Lambda$.
Practically, we use the PU bound by replacing the coupling constants in Eqs.~\eqref{eq:PU-scalar1}-\eqref{eq:PU-scalar2} and \eqref{eq:PU-gauge} with the running coupling constants, namely $\lambda_{1} (\mu)$, $\lambda_{3} (\mu)$, and $g_V(\mu)$, and define $\Lambda$ as a scale at which the inequalities are violated.
We take initial conditions of the RG equations at $\mu = m_\text{DM}^{}$ as to be the coupling constants calculated from Eqs.~\eqref{eq:parameter-pot-1}-\eqref{eq:parameter-pot-2}, and \eqref{eq:Z'mass}.
The white region indicates
$\Lambda>10^{18}$ GeV, 
which means that they do not hit the Landau pole until the Planck scale,
while the orange hatched and  gray regions indicate
$\Lambda<100$ TeV and $100\, \mr{TeV}<\Lambda<10^{18} \, \mr{GeV}$, respectively.
In the top, middle, and bottom panels, the blue thick, dashed orange, and dotted red curves indicate the parameter space providing the correct DM abundance
and corresponds to the curves shown in Fig.~\ref{fig:comparison}.
The parameters are taken as $m_{h'}=300\, \mr{GeV},~ m_{s_-}=1.5 \,m_\mr{DM}^{}\,~ \sin \epsilon=10^{-4},~ \sin \xi=0.1$,
and $m_{Z'}^{}=200\, \mr{GeV}$ (top panel), $m_{Z'}^{}=3 \, m_\mr{DM}^{}$ (middle panel), or $m_{Z'}^{}=1.1 \, m_\mr{DM}^{}$ (bottom panel).
In the case of the middle panel, in most parameter space, the model breaks down and requires some UV completion at the cutoff scale around $100$ TeV, which resembles the behavior of the model in Ref.~\cite{Abe:2022mlc}.
On the other hand, in the case of the top panel, the model basically remains valid up to $100$ TeV, and in some parameter space as high as the Planck scale.
This is a sharp contrast to the model in Ref.~\cite{Abe:2022mlc}, and the benefit coming from the exchange symmetry introduced in this model.
In the bottom panel, the constraint by the Landau pole is slightly milder than the middle one
while it still requires some UV completion by $100$ TeV for $m_\mr{DM}\gtrsim 800$ GeV.

Let us give a comment on a case that $s_-$ has an approximately degenerated mass with the pNG DM $a$.
In this case, co-annihilation processes, $a s_- \to Z'^\ast (Z^\ast) \to f\bar f$, can happen.
In addition, the decay process $s_- \to a Z'^\ast$ is suppressed and $s_-$ can be long-lived.
As stated in Sec.~\ref{sec:improved-model}, such a degeneracy is realized if and only if $\lambda_1\simeq \lambda_3$,
which enlarges the symmetries in the scalar potential to the global $O(4)$ symmetry as presented in Appendix.~\ref{sec:appendix},
and hence the model reduces to the one studied in Ref.~\cite{Abe:2022mlc},
in which the pNG DM is given as a complex scalar $s_- + i a$.

\clearpage
\section{Loop induced spin-independent cross section}\label{sec:loop}

The DM direct detection experiments utilize a scattering of DM off a nucleon. The incoming particles are non-relativistic, and thus, the momentum transfer is quite small. As discussed in Sec.~\ref{sec:DDatTree}, the amplitude of a pNG DM particle scattering off a nucleon is highly suppressed by the small momentum transfer, and thus the spin-independent cross section $\sigma_\text{SI}^{}$ is essentially zero at the tree level analysis.
This suppression is due to a property of NG bosons. 
However, the DM candidate 
is not a NG boson but \textit{a pseudo-NG boson} due to the explicit global $U(1)$ breaking by $m_{12}^2$. 
Hence, the suppression is not guaranteed at the loop level in general. 
In fact, the spin-independent cross section is induced at the loop level in other pNG models\footnote{Loop effects in models similar to pNG DM models were studied in~\cite{Cho:2023hek}.}~\cite{Ishiwata:2018sdi, Azevedo:2018exj, Glaus:2020ihj, Abe:2022mlc}. In this section, we investigate the loop effect on the spin-independent cross section in our model.


If 
the DM candidate were massless and a NG boson, the scattering amplitude would be suppressed even at the loop level. 
This ensures that
the loop-level scattering amplitude vanishes in the limit of $m_{12}^2 \to 0$
in which
the DM candidate becomes a massless NG boson,
namely,
\begin{equation}
    \lim_{m_{12}^2 \to 0} \sum_\text{diagrams}\mathcal{M}(m_{12}^2) = 0.
\end{equation}
Using this property, we can simplify our loop calculation~\cite{Abe:2022mlc}.
The total scattering amplitude can be rewritten as
\begin{equation}
\sum_\text{diagrams}\mathcal{M}(m_{12}^2) = \sum_\text{diagrams}\left(\mathcal{M}(m_{12}^2) - \mathcal{M}(0) \right) \, ,
\end{equation}
from which it follow that 
the diagrams independent of $m_{12}^2$ cancel within the parenthesis and do not contribute to the final expression of the scattering amplitude. 
We calculate only the diagrams that have $m_{12}^2$ dependence, and subtract $m_{12}^2$ independent part.
We find that $m_{12}^2$ dependence appears in $m_{a}$, $m_{s_{-}}$, the vertices including $s_{-}$, and the $a a \pi_{V}^{} \pi_{V}^{}$-coupling, where $\pi_V^{}$ is the would-be NG boson that is eaten by $Z'$.\footnote{In this section, we ignore the kinetic mixing because its effect is negligible.}
The expressions of the amplitude for each diagram are given in Appendix~\ref{app:loop}.

Figure~\ref{fig:xsec-loop} shows the spin-independent cross section $\sigma_\text{SI}^{}$ at the loop level.
Here, we take the same parameter set as in Fig.~\ref{fig:comparison}, but with vanishing kinetic mixing, namely $\epsilon = 0$. 
We find that the loop-induced values of $\sigma_\text{SI}^{}$ typically drop into the neutrino fog. In some regions, $\sigma_\text{SI}^{}$ are not hidden by the neutrino fog. 
However, the cutoff scale $\Lambda$ estimated from the RG running is below $100$~TeV in those regions, see also Fig.~\ref{fig:Landau-pole}. 
In the viewpoint of effective field theories, higher dimensional operators can exist in the Lagrangian with coefficients inversely proportional to several powers of $\Lambda$. Such operators modify our prediction based on the renormalizable model. 
Thus, the low cutoff scale is disfavored from the viewpoint of model predictability and requires some UV completion.
As a result, in order to probe the plausible parameter space of the model, it is necessary to probe the region in the neutrino fog.
%
\begin{figure}[tbp]
\centering
\includegraphics[width=0.55\textwidth]{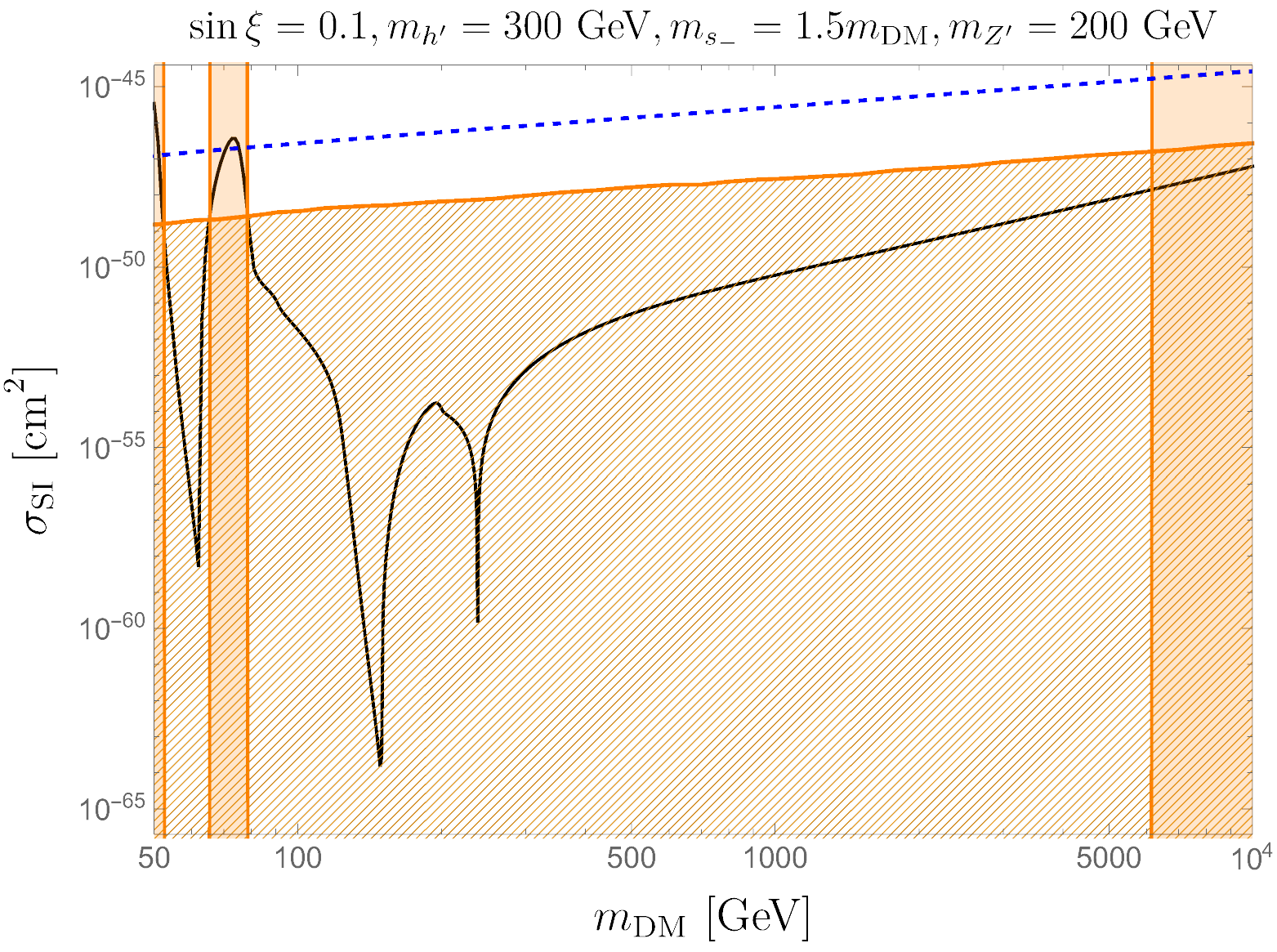}\\[5ex]
\includegraphics[width=0.55\textwidth]{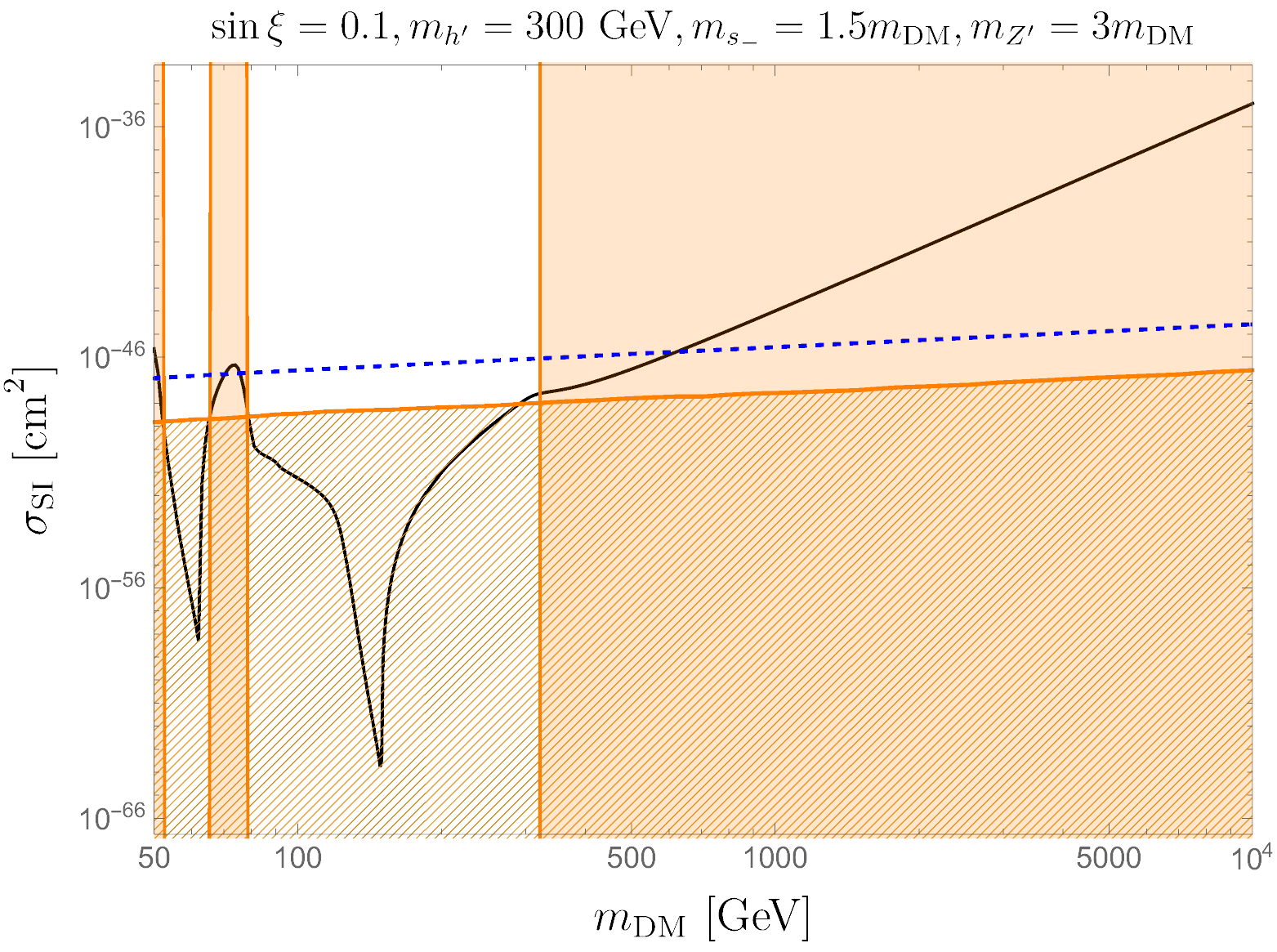}\\[5ex]
\includegraphics[width=0.55\textwidth]{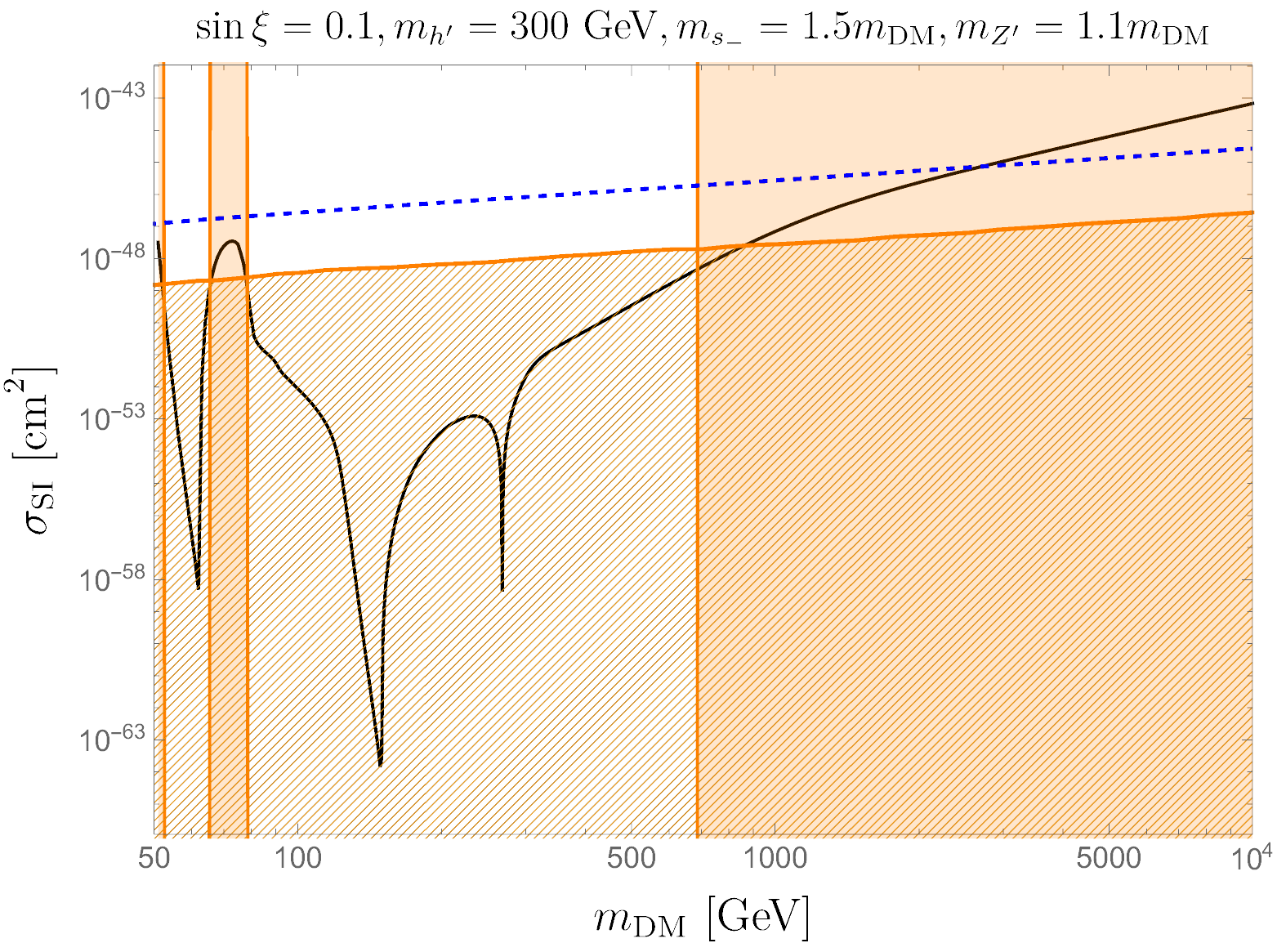}
\caption{
The loop induced spin-independent cross section. 
The top, middle, and bottom panels are for $m_{Z'}^{} = 200$~GeV, $m_{Z'}^{} = 3\, m_\text{DM}^{}$, and $m_{Z'}^{} = 1.1\,  m_\text{DM}^{}$, respectively.
The black-thick curve is the model prediction.
The current most stringent bound is given by the LZ experiment~\cite{LUX2022} and is denoted by the blue-dashed line. 
The region of the neutrino fog is filled by the orange hatched pattern.
In the orange-filled regions, the cutoff scale estimated from the RG evolution is less than 100~TeV. 
}
\label{fig:xsec-loop}
\end{figure}

\clearpage
\section{Discussion and conclusion}
\label{sec:conclusion}

We have proposed the new pNG DM model with two SM-singlet complex scalars and a dark $U(1)$ gauge field.
The model also has a global $U(1)$ symmetry and a $\mathbb{Z}_2$ symmetry under the exchange of the two scalars.
The global $U(1)$ symmetry is broken explicitly and softly by the mass-dimension-two term,
where other soft-breaking terms are forbidden by the gauge $U(1)$ symmetry.
The VEVs of the scalars spontaneously break the both $U(1)$ symmetries, 
giving rise to the pNG boson whose mass is originated from the soft-breaking parameter.
The exchange symmetry ensures the stability of the pNG DM $a$ as well as the suppression of the DM-nucleon scattering cross section. 
This model does not suffer from the DW problem.
The comparison with the other pNG DM models are shown in Tab.~\ref{tab:table-models}. 
We have discussed the phenomenology of this model and shown that it can explain the correct DM relic abundance at present universe within the theoretical and experimental constraints.
Note that if $m_{Z'}^{}$ is lighter than $m_\mr{DM}^{}$,
the annihilation process $aa\to Z'Z'$ is open and gives the relatively large annihilation cross section.
Consequently, the parameter dependence of the DM relic abundance on $v_s$ and $m_\mr{DM}^{}$ can be significantly different in heavier $m_\mr{DM}^{}$ region from those in the other pNG DM models, as shown in Fig.~\ref{fig:relic} and Fig.~\ref{fig:comparison}.
This leads to the relatively large VEV and small scalar and gauge coupling constants,
and hence, the Landau pole can be evaded until the high-energy scale as shown in Fig.~\ref{fig:Landau-pole}.
We have also calculated the loop-induced spin-independent cross section $\sigma_\mr{SI}^{}$ with nucleons as shown in Fig.~\ref{fig:xsec-loop}.
We found that the favorable parameter space of the model is basically covered by the neutrino fog.

We here comment on topological solitons appearing in this model.
Because the vacuum manifold in the dark sector has the non-trivial first homotopy group,
the model admits topologically stable vortex strings.
When $m_{h'}^{}$ is heavier than $m_{Z'}^{}$, the cross section of the vortex solution is not axially symmetric but has a dipole-like structure consisting of two half-quantized vortices connected by a domain wall,
as studied in Ref.~\cite{Eto:2016mqc} in the semilocal model~\cite{Vachaspati:1991dz,Achucarro:1999it} with an explicit breaking term.
This non-trivial structure of the strings could lead to their non-trivial dynamics and help us to probe this model in cosmological and astrophysical observations.
The detailed studies will be reported elsewhere.

\section*{Acknowledgments}
This work is supported in part by the MEXT Grant-in-Aid for Scientific Research on Innovation Areas Grant No. JP18H05543 (K.T.)
and JSPS KAKENHI Grant Numbers 
21K03549 (T.A.), JP21J01117 (Y.H.) and JP22K03620 (K.T.). 
The work is also supported by JSPS Core-to-Core Program (grant number:JPJSCCA20200002)
and the Deutsche Forschungsgemeinschaft under Germany's Excellence Strategy - EXC 2121 Quantum Universe - 390833306.


\appendix


\section{The model with $O(4)$ symmetry}
\label{sec:appendix}
When one takes $\lambda_1=\lambda_3$ in the model introduced in Sec.~\ref{sec:improved-model}, 
this model enjoys a (softly-broken) global $O(4)$ symmetry.
Then $s_-$ and $a$ are degenerate in mass and form a complex pNG boson of the $O(4)$ symmetry breaking.
This is made obvious in the following basis:
\begin{align}
 \phi_+ &= \frac{1}{\sqrt{2}}\left(\phi_1+ \phi_2\right) =  \frac{1}{\sqrt{2}}\left(v_s + s_+ + i \pi_V^{}\right) \, , \\
 \phi_- & = \frac{1}{\sqrt{2}}\left(\phi_1- \phi_2\right)=  \frac{1}{\sqrt{2}}\left(s_- + i a\right)  \, .
\end{align}
Using $\lambda_1=\lambda_3$ and 
\begin{equation}
 |\phi_1|^2 + |\phi_2|^2 = |\phi_+|^2 + |\phi_-|^2 \, ,
\end{equation}
the potential in this new basis is written as 
\begin{align}
V (H,\phi_+ , \phi_-) & = m_1^2 (|\phi_+|^2 + |\phi_-|^2)  - m_{12}^2 (|\phi_+|^2 - |\phi_-|^2) \n \\
&\hspace{3em} + \f{\lambda_1}{2} \left(|\phi_+|^2 + |\phi_-|^2\right)^2 \n \\
&\hspace{3em} - m_H^2 |H|^2 + \lambda_H |H|^4 + \lambda_{H1} |H|^2 \left(|\phi_+|^2 + |\phi_-|^2 \right) \, .
\end{align}
It is found that $a$, $\pi_V^{}$, and $s_-$ are (p)NG bosons of this $O(4)$ symmetry.
While $\pi_V^{}$ is eaten by the gauge field, the others get their masses $2m_{12}^2$ due to the explicit breaking term proportional to $m_{12}^2$.

If one introduces a complex doublet $\bm{\phi}$,
\begin{equation}
 \bm{\phi} \equiv
\begin{pmatrix}
 \phi_- \\ \phi_+
\end{pmatrix} \, ,
\end{equation}
then the potential is rewritten as
\begin{align}
V (H,\bm{\phi}) & = m_1^2 |\bm{\phi}|^2  + m_{12}^2 \bm{\phi}^\dagger \sigma_3 \bm{\phi} + \f{\lambda_1}{2} \left(|\bm{\phi}|^2 \right)^2 \n \\ 
&\hspace{1em} - m_H^2 |H|^2 + \lambda_H |H|^4 + \lambda_{H1} |H|^2 |\bm{\phi}|^2 \, ,
\end{align}
which is invariant under the gauge transformation $U(1)_V$
\begin{equation}
 \bm{\phi} \to e^{i\alpha(x)} \bm{\phi} \, ,
\end{equation}
and hence this model is equivalent to that in Ref.~\cite{Abe:2022mlc}.
In this case, the two real components $s_-$ and $a$ form a complex DM, which is denoted by $\chi$ in Ref.~\cite{Abe:2022mlc}.
Therefore, the model introduced in Sec.~\ref{sec:improved-model} can be regarded as a model obtained by decomposing the complex DM $\chi$ into real and imaginary parts and splitting their masses.

\section{The model without exchange symmetry}
\label{sec:U1xU1model}

We here discuss a model generalized by removing the exchange symmetry from the model discussed above.
The field contents are the same as those.
\subsection{Lagrangian}
Without the exchange symmetry between $\phi_1$ and $\phi_2$,
the Lagrangian is given as
\begin{equation}
 \mathcal{L}=  \mathcal{L}_\text{\{SM w/o Higgs potential\}} +|D_\mu \phi_1|^2 + |D_\mu \phi_2|^2 - \frac{1}{4}V^{\mu\nu}V_{\mu\nu} - \frac{\sin\epsilon}{2}V^{\mu\nu}Y_{\mu\nu} - V(H,\phi_1 , \phi_2)
\end{equation}
with the scalar potential
\begin{align}
V (H,\phi_1 , \phi_2) & = m_1^2 |\phi_1|^2 +m_2^2 |\phi_2|^2  - \left(m_{12}^2 \phi_1^\ast \phi_2 + \mr{h.c.}\right) \n  \\
&\hspace{3em} + \f{\lambda_1}{2} |\phi_1|^4 + \f{\lambda_2}{2}|\phi_2|^4 + \lambda_3 |\phi_1|^2 |\phi_2|^2 \n \\
&\hspace{3em} - m_H^2 |H|^2 + \lambda_H |H|^4 + \lambda_{H1} |H|^2 |\phi_1|^2 + \lambda_{H2} |H|^2 |\phi_2|^2 \, .
\end{align}

In general, $\phi_1$ and $\phi_2$ take different VEVs as follows,
\begin{align}
 \langle H \rangle = \f{1}{\sqrt{2}}
\begin{pmatrix}
 0 \\ v 
\end{pmatrix}, \quad
\langle\phi_1\rangle = \f{v_1}{\sqrt{2}} ,\quad
\langle\phi_2\rangle = \f{v_2}{\sqrt{2}} ,
\end{align}
The four real scalar particles around the vacuum are given as
\begin{equation}
 H = \f{1}{\sqrt{2}}
\begin{pmatrix}
 0 \\ v + \sigma(x)
\end{pmatrix}\, , 
\end{equation}
\begin{equation}
\phi_1 = \f{1}{\sqrt{2}} \left(v_1 + s_1(x) + i a (x)\sin \beta \right), \quad
\phi_2 = \f{1}{\sqrt{2}} \left(v_2 + s_2(x) - i a(x)\cos \beta \right) \, ,
\end{equation}
with $\sin \beta \equiv v_2 / v_s$, $\cos \beta \equiv v_1 /v_s$ and $v_s\equiv\sqrt{v_1^2 +v_2^2}$.
Here one can take $v_1$ and $v_2$ as real without loss of generality by redefinition of the fields.
The fields $\sigma$, $s_1$, and $s_2$ are CP-even real scalars while $a$ is a CP-odd real scalar. 
We have taken the unitary gauge for the $U(1)_V$ gauge and the SM gauge symmetries and have taken $a$ as the direction orthogonal to $U(1)_V$.

\subsection{Scalar mass spectrum}

Substituting the fields by the components in the potential, we obtain the stationary conditions
\begin{align}
\begin{cases}
m_H^2 = \frac{1}{2} \left(\lambda_{H1} v_1^2 + \lambda_{H2} v_2^2 \right)+ \lambda_H v^2 \\
\mathrm{Im} \,m_{12}^2 =0 \\
m_1^2 = \mathrm{Re}\, m_{12}^2 \tan \beta - \f{1}{2}\lambda_{H1} v^2 - \f{1}{2}\lambda_1 v_1^2 - \f{1}{2}\lambda_3  v_2^2 \\
m_2^2 = \mathrm{Re}\, m_{12}^2 \f{1}{\tan \beta} - \f{1}{2}\lambda_{H2} v^2 - \f{1}{2}\lambda_2 v_2^2 - \f{1}{2}\lambda_3  v_1^2 ,
\end{cases}
\end{align}
from which it follows that $m_{12}^2$ should be real.
Then we have the following mass terms
\begin{align}
 \mathcal{L} \supset 
- \f{1}{2}
\begin{pmatrix}
 \sigma & s_1 & s_2
\end{pmatrix}
M^2_\mr{even}
\begin{pmatrix}
 \sigma \\ s_1 \\ s_2
\end{pmatrix}
 - \f{1}{2} M_a^2 a^2
\end{align}
with
\begin{equation}
 M^2_\mr{even} = 
 \begin{pmatrix}
  2\lambda v^2 & \lambda_{H1} v v_1 &  \lambda_{H2} v v_2 \\
   \lambda_{H1} v v_1 & \f{m_{12}^2 v_2}{v_1} + \lambda_1 v_1^2 & \lambda_3 v_1 v_2 - m_{12}^2 \\
   \lambda_{H2} v v_2 & \lambda_3 v_1 v_2 - m_{12}^2 & \f{m_{12}^2 v_1}{v_2} + \lambda_2 v_2^2 \\
 \end{pmatrix}
\end{equation}
and
\begin{equation}
 M_a^2 = \f{m_{12}^2 }{\sin \beta \cos \beta} \equiv m_\mathrm{DM}^2.
\end{equation}
We can move on to the mass eigenstates for the neutral CP-even scalars as
\begin{align}
\begin{pmatrix}
 \sigma \\ s_1 \\ s_2
\end{pmatrix}
&= R
\begin{pmatrix}
 h_{125} \\ \eta_1 \\ \eta_2
\end{pmatrix} 
\end{align}
with the $O(3)$ matrix $R$ being described by the Euler angles as
\begin{align}
R &=
\begin{pmatrix}
1 & 0 & 0 \\
0 & \cos\theta_2 & -\sin \theta_2 \\
0 & \sin \theta_2 & \cos \theta_2
\end{pmatrix}
\begin{pmatrix}
\cos \theta_1 & 0 & \sin \theta_1 \\
0 & 1 &  0 \\
-\sin \theta_1 & 0 & \cos \theta_1 \\
\end{pmatrix}
\begin{pmatrix}
1 & 0 & 0 \\
0 & \cos\theta_3 & -\sin \theta_3 \\
0 & \sin \theta_3 & \cos \theta_3
\end{pmatrix} \, ,
\end{align}
which is supposed to diagonlalize the mass matrix as
\begin{equation}
R^T M^2_\mr{even} R = 
\mathrm{diag} ~
(m_h^2, m_{\eta_1}^2, m_{\eta_2}^2) \,.
\end{equation}

\subsection{Stability of pNG DM}
When the gauge kinetic mixing parameter $\epsilon$ is non-zero,
the pNG DM $a$ can decay via processes with a vertex $a V_\mu s_{1(2)}$
where the off-shell gauge boson $V_\mu$ and the scalar $s_{1(2)}$ eventually decay into SM light particles,
resulting in that $a$ cannot be a stable DM.
To avoid this in the generalized model, we must impose a discrete symmetry under the dark CP conjugation
\begin{align}
\begin{cases}
    V_\mu \to  - V_\mu \\
    \phi_1 \to \phi_1^* \\
    \phi_2 \to \phi_2^*
\end{cases}  
\end{align}
which leads to
\begin{equation}
    \sin\epsilon=0 \, .
\end{equation}
Thanks to this condition, the pNG $a$ and $V_\mu$ have odd charges under this $\mathbb{Z}_2$ transformation while the others having even charges,
and hence they cannot decay into SM particles.
In addition, by imposing the mass of $V_\mu$ to be heavier than $a$, 
the decay of $a$ such as $a \to V_\mu f \bar f $ is kinematically forbidden,
making $a$ stable.

\subsection{Non-suppression of scattering cross section in direct detection}
The most plausible feature of the original pNG DM model~\cite{Gross:2017dan} is that the spin-independent scattering cross section of the pNG DM with nucleons vanishes at the tree level with the zero momentum transfer.
This is due to, in the linear representation for the pNG boson, the cancellation between two Feynman diagrams mediated by two neutral scalar particles.
This is also the case for similar models, such as models in Refs.~\cite{Abe:2021byq,Okada:2021qmi,Liu:2022evb,Abe:2022mlc,Otsuka:2022zdy},
in which the pNG boson obtains its mass from explicit breaking terms with mass dimension two. 
On the other hand, in other models containing explicit breaking terms with mass dimension one or three~\cite{Abe:2021nih},
the cancellation does not work, leading to the sizable spin-independent cross section, unless the masses of the mediating scalars are tuned to be degenerated.

Therefore, one may consider that the present model works well as a pNG DM model since the explicit breaking term is given 
as the operator with mass dimension two, $\phi_1^* \phi_2 + \mr{h.c.}$
However, this is not the case.
We here show that the spin-independent cross section of the pNG DM $a$ with nucleons is not suppressed in general.
To see this, it is convenient to move on to the non-linear representation, 
\begin{align}
 \phi_1 &= \frac{1}{\sqrt{2}} (v_1 +s_1) \exp\left[i \sin \beta \frac{\pi_a}{v_1}\right] \, ,\\
 \phi_2 &= \frac{1}{\sqrt{2}} (v_2 +s_2) \exp\left[-i \cos \beta \frac{\pi_a}{v_2}\right]  \, ,
\end{align}
in which $\pi_a$ corresponds to $a$ in the linear representation and we have taken the unitary gauge again.
In this representation, the vanishment of the cross section is caused by the vanishment of the cubic couplings $\pi_a \pi_a s_1$ and $\pi_a \pi_a s_2$ instead of non-trivial cancellation between different diagrams, as is studied in Ref.~\cite{Cai:2021evx}.

To read off the cubic couplings, 
we substitute these expressions into the Lagrangian and obtain
%
%
%
\begin{align}
 & |\partial_\mu \phi_1|^2 +  |\partial_\mu \phi_2|^2 + m_{12}^2 \left(\phi_1^\ast \phi_2 + \mathrm{h.c.}  \right)\\
& \supset 
\frac{1}{2 } \left[(v_1 +s_1)^2 \tan^2 \beta + (v_2 +s_2)^2 \frac{1}{\tan^2 \beta} \right]\left(\frac{\partial_\mu \pi_a}{v_s}\right)^2
 \nonumber \\
& \hspace{3em}-  
\frac{1}{2} m_\mathrm{DM}^2 (v_1 + s_1)(v_2 + s_2) \frac{1}{\sin\beta\cos \beta}\left(\frac{\pi_a}{v_s}\right)^2 \, ,
\end{align}
from which we get the cubic couplings,
\begin{align}
\mathcal{L} &\supset
\frac{m_\mathrm{DM}^2 }{4v_s} (\sin2\beta+\cos2\beta + 1)(1-\tan\beta) \left(-\frac{s_1}{\cos\beta}+\frac{s_2}{\sin\beta}\right) \pi_a^2\,, \label{eq:aas-general}
\end{align}
where we have used that the momenta carried by $s_1$ and $s_2$ are negligible in the direct detection experiments and the on-shell condition for $\pi_a$, $(\partial_\mu \partial^\mu + m_\mathrm{DM}^2) \pi_a=0$.
Therefore, these couplings between the pNG $\pi_a$ and the scalars $s_1$ and $s_2$ do not vanish in general,
leading to a sizable spin-independent cross section with nucleons.
This is not necessarily inconsistent with the low energy theorem (soft pion theorem) since the above couplings are proportional to $m_\mr{DM}^{}$ and hence vanish in the limit $m_\mr{DM}^{}\to 0$, where the pNG DM $\pi_a$ becomes a true NG boson.
From the above expression, it is obvious that the vanishment of the couplings (and hence the cross section) requires a tuning of $\tan\beta \to 1$.
In the model introduced in Sec.~\ref{sec:improved-model} with the exchange symmetry,
$\tan\beta=1$ is naturally realized, and hence the amplitude vanishes at the tree level as shown.

\section{Derivation of boundedness condition of potential}
\label{sec:deriv-boundedness}
We here derive necessary and sufficient conditions \eqref{eq:boundedness} that the scalar potential is bounded from below.
Since they are irrelevant for the large field values, we ignore all quadratic terms in the potential and concentrate on the quartic terms:
\begin{align}
  V_\mr{quart.}&=\lambda_{H} |H|^4 + \frac{\lambda_1}{2}\left(|\phi_1|^4 + |\phi_2|^4\right) + \lambda_3 |\phi_1|^2 |\phi_2|^2 
 + \lambda_{H1} |H|^2 (|\phi_1|^2 + |\phi_2|^2) \, .
\end{align}
Because of the exchange symmetry in the quartic terms, we can take $|\phi_1|\geq|\phi_2|$ without loss of generality.

\paragraph{Case of $\lambda_{H1}>0$}
Remark that the potential is rewritten as
\begin{align}
 V_\mr{quart.}&=\lambda_H \left( |H|^2 - \sqrt{\frac{\lambda_1}{2\lambda_H}} \left(|\phi_1|^2-|\phi_2|^2\right) \right)^2 \nonumber \\[1ex]
&\hspace{1em}+ \left(\lambda_{H1} + \sqrt{2\lambda_H \lambda_1}\right) |H|^2 \left(|\phi_1|^2-|\phi_2|^2\right)\nonumber \\[1ex]
& \hspace{1em} + 2 \lambda_{H1} |H|^2 |\phi_2|^2 + \left(\lambda_1+\lambda_3\right) |\phi_1|^2| | \phi_2|^2 \, .\label{eq:bounded-pot1}
\end{align}
We first obtain necessary conditions for $V_\mr{quart.}$ to be non-negative with arbitrary directions as follows.
For a direction in which $|\phi_1|=|\phi_2|$ and $|H|=0$ in \eqref{eq:bounded-pot1},
we have
\begin{align}
 V_\mr{quart.} = (\lambda_1+\lambda_3)  |\phi_1|^2 |\phi_2|^2  \quad\Rightarrow \lambda_1 + \lambda_3 >0 \, .
\end{align}
For a direction in which $|\phi_2|=0$ and $|H|=0$ in \eqref{eq:bounded-pot1},
we have
\begin{align}
 V_\mr{quart.} = \frac{\lambda_1}{2}  |\phi_1|^4  \quad\Rightarrow \lambda_1 >0 \, .
\end{align}
For a direction in which $|\phi_1|=0$ and $|\phi_2|=0$ in \eqref{eq:bounded-pot1},
we have
\begin{align}
 V_\mr{quart.} = \lambda_H |H|^4  \quad\Rightarrow \lambda_H >0 \, .
\end{align}
Thus, in the case of $\lambda_{H1}>0$, we summarize the necessary conditions as
\begin{align}
 \lambda_1 + \lambda_3 >0, \quad \lambda_1 >0, \quad \lambda_H >0 \, .\label{eq:bounded-condition1}
\end{align}

We can check these are also sufficient conditions as follows.
Assuming these and $|\phi_1|\geq|\phi_2|$, it is obvious that all terms in the potential \eqref{eq:bounded-pot1} are positive
\begin{align}
 V_\mr{quart.} &\geq 0,
\end{align}
where the equality holds when
\begin{equation}
 |H|^2=\sqrt{\frac{\lambda_1}{2\lambda_H}} \left(|\phi_1|^2-|\phi_2|^2\right) \, ,
\quad |\phi_1|=|\phi_2|=0 \, .
\end{equation}
(We have also used $\lambda_{H1}>0$.)
Therefore we have found \eqref{eq:bounded-condition1} to be necessary and sufficient conditions for the case of $\lambda_{H1}>0$.

\paragraph{Case of $\lambda_{H1}<0$} 
We rewrite the potential $V_\mr{quart.}$ as
\begin{align}
 V_\mr{quart.}
%
&=\lambda_{H} \left[|H|^2 + \frac{\lambda_{H1}}{2 \lambda_H}\left(|\phi_1|^2 + |\phi_2|^2\right)\right]^2 \nonumber \\[1ex]
&\hspace{1em}+ \left(\frac{\lambda_1}{2}-\frac{\lambda_{H1}^2}{4 \lambda_H} \right) \left(|\phi_1|^2 -  |\phi_2|^2\right)^2
+ \left[\lambda_3 + \lambda_1 - \frac{\lambda_{H1}^2}{\lambda_H} \right]|\phi_1|^2 |\phi_2|^2  \label{eq:bounded-pot2} \, .
\end{align}
Again, we obtain necessary conditions for $V_\mr{quart.}$ to be non-negative with arbitrary directions as follows.
For a direction in which $|\phi_1|=|\phi_2|=0$  in \eqref{eq:bounded-pot2},
we have
\begin{align}
 V_\mr{quart.} = \lambda_H |H|^4  \quad\Rightarrow \lambda_H >0 \, .
\end{align}
For a direction in which $|H|^2 = -\frac{\lambda_{H1}}{\lambda_H}(|\phi_1|^2 + |\phi_2|^2)$ and $|\phi_1|=|\phi_2|$,
we have
\begin{align}
 V_\mr{quart.} &= \left[\lambda_3 + \lambda_1 - \frac{\lambda_{H1}^2}{\lambda_H} \right]|\phi_1|^2 |\phi_2|^2  \\[1ex]
&\Rightarrow \lambda_3 + \lambda_1 - \frac{\lambda_{H1}^2}{\lambda_H}>0 \\[1ex]
& \Rightarrow -\sqrt{(\lambda_1 + \lambda_3)\lambda_H} < \lambda_{H1} < 0 , \quad \lambda_1+\lambda_3>0\, .
\end{align}
For a direction in which $|H|^2 =- \frac{\lambda_{H1}}{\lambda_H}(|\phi_1|^2 + | \phi_2|^2)$ and $|\phi_2|=0$,
we have
\begin{align}
 V_\mr{quart.} &= \left(\frac{\lambda_1}{2}-\frac{\lambda_{H1}^2}{4 \lambda_H} \right) \left(|\phi_1|^2 \right)^2 \\[1ex]
&\Rightarrow \frac{\lambda_1}{2}-\frac{\lambda_{H1}^2}{4 \lambda_H}>0 \\[1ex]
& \Rightarrow -\sqrt{2\lambda_1 \lambda_H} < \lambda_{H1} < 0 , \quad  \lambda_1>0 \, .
\end{align}
Thus, in the case of $\lambda_{H1}<0$, we summarize the necessary conditions as
\begin{align}
 \lambda_1 + \lambda_3 >0, \quad \lambda_1 >0, \quad \lambda_H >0,
\quad \lambda_{H1} + \sqrt{2\lambda_H \lambda_1} >0 , 
\quad \lambda_{H1} + \sqrt{(\lambda_1 + \lambda_3)\lambda_H} >0 \, .
\label{eq:bounded-condition2}
\end{align}

Again, it is obvious that these conditions are sufficient conditions 
because assuming them, all terms in the potential \eqref{eq:bounded-pot2} are non-negative.
Therefore \eqref{eq:bounded-condition2} are necessary and sufficient conditions for $\lambda_{H1}<0$.

Since the conditions $\lambda_{H1} + \sqrt{2\lambda_H \lambda_1} >0$ and
$\lambda_{H1} + \sqrt{(\lambda_1 + \lambda_3)\lambda_H} >0$ are automatically satisfied for $\lambda_{H1}>0$,
we can combine the two cases of $\lambda_{H1}>0$ and $\lambda_{H1}<0$
to obtain the necessary and sufficient conditions \eqref{eq:boundedness}.


\section{Details of the Loop calculation for $\sigma_\text{SI}^{}$}\label{app:loop}
In this section, we calculate the diagrams that are relevant for the DM-nucleon scattering process. 
We calculate all the diagrams with zero-momentum transfer. 
Note that the gauge kinetic mixing is negligible.
In the following, we denote $h$ ($h'$) as $h_1$ ($h_2$), for simplicity.
We perform our calculation in the Landau gauge although the result is independent of the gauge choice.

\begin{figure}[tbp]
\centering
\includegraphics[width=0.25\textwidth]{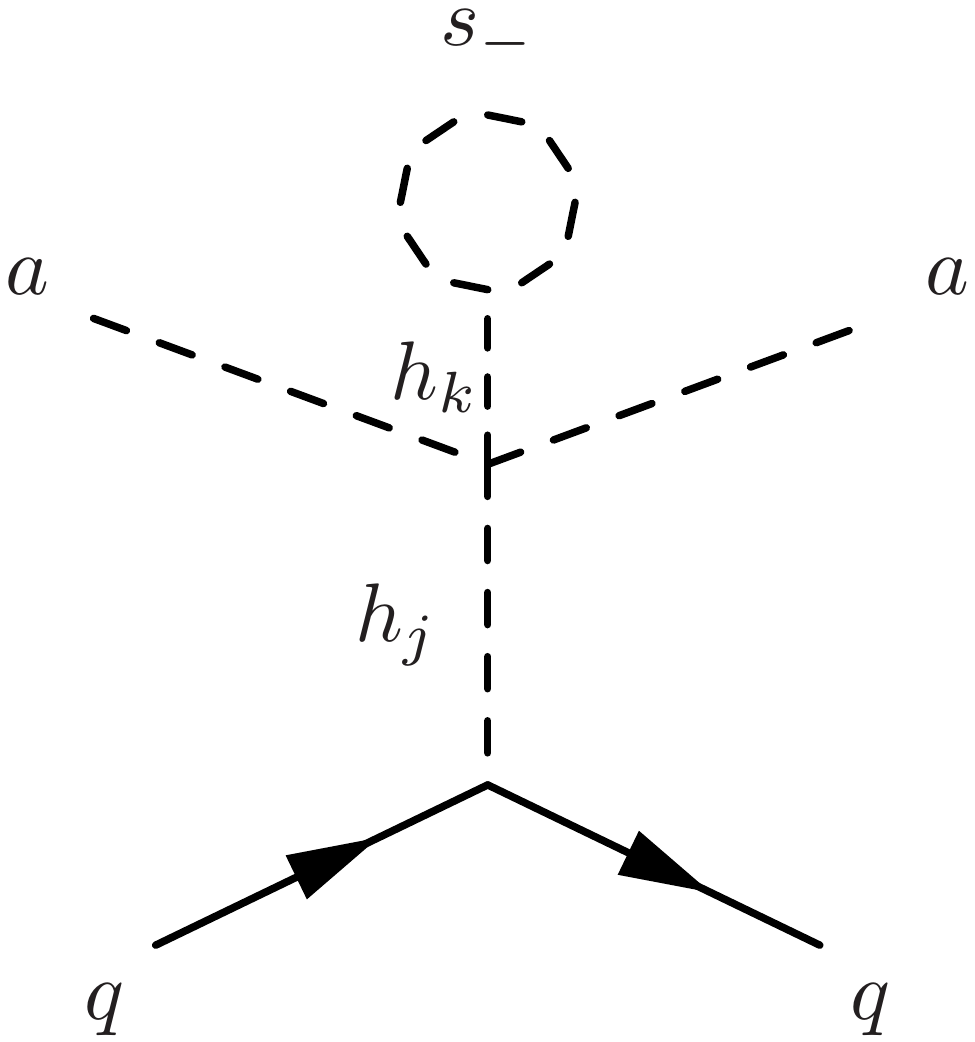}
\includegraphics[width=0.25\textwidth]{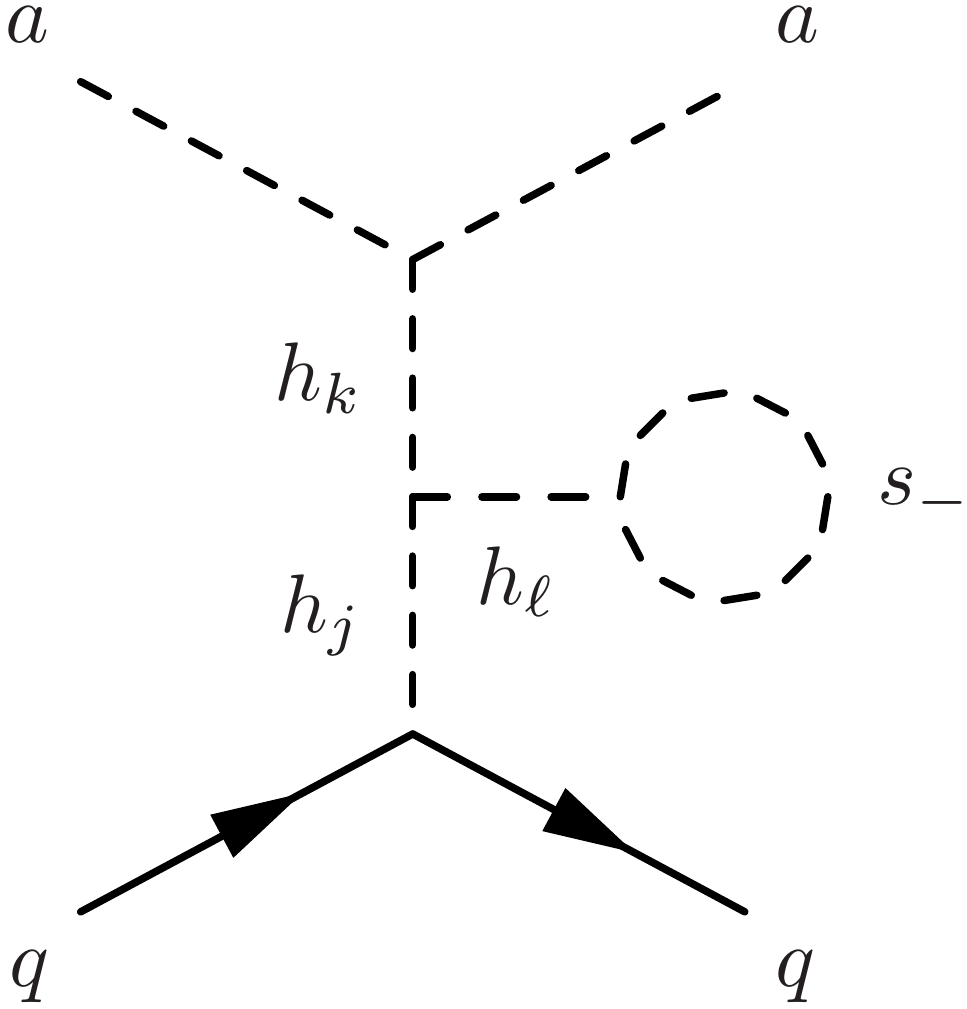}
\includegraphics[width=0.25\textwidth]{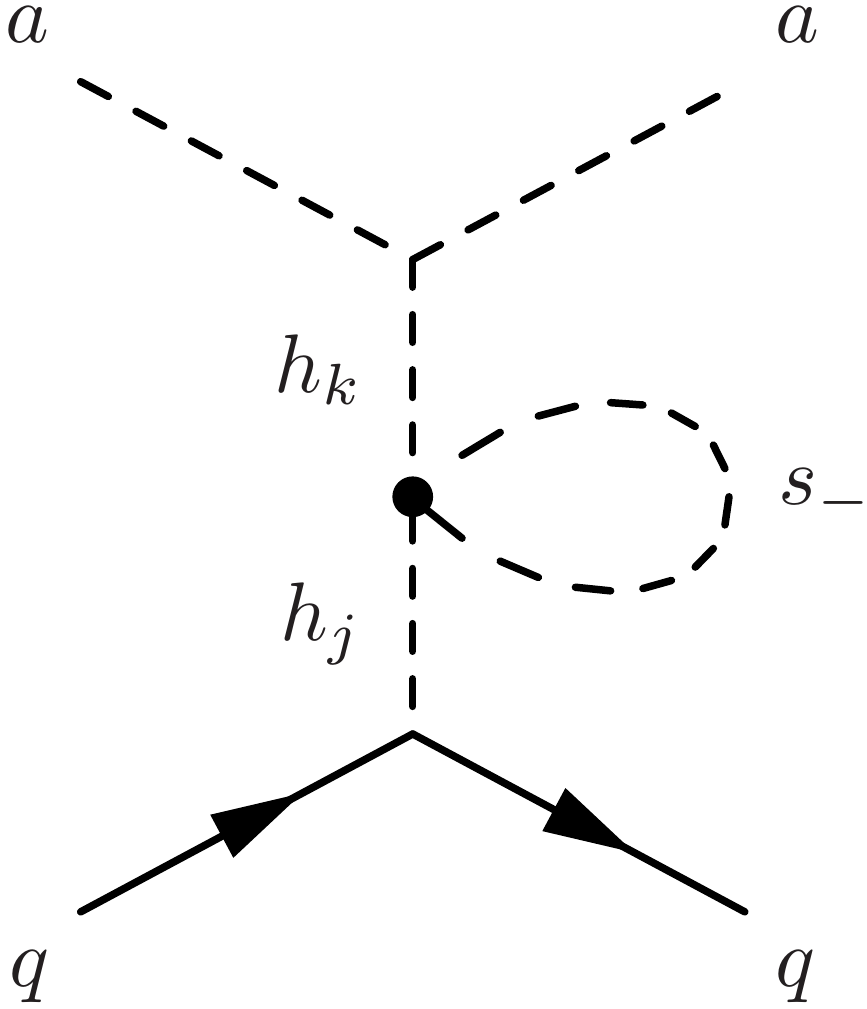}
\ \\ \ \\ \ \\ 
\includegraphics[width=0.25\textwidth]{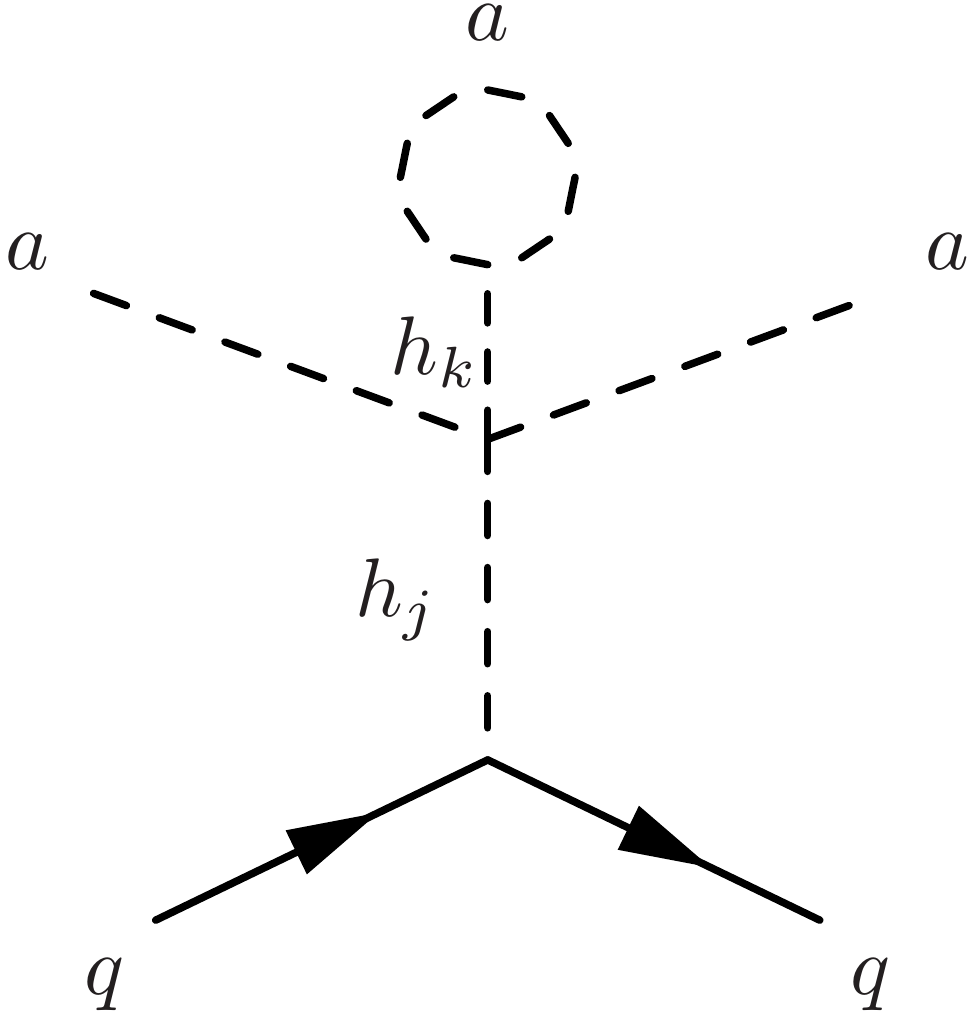}
\includegraphics[width=0.25\textwidth]{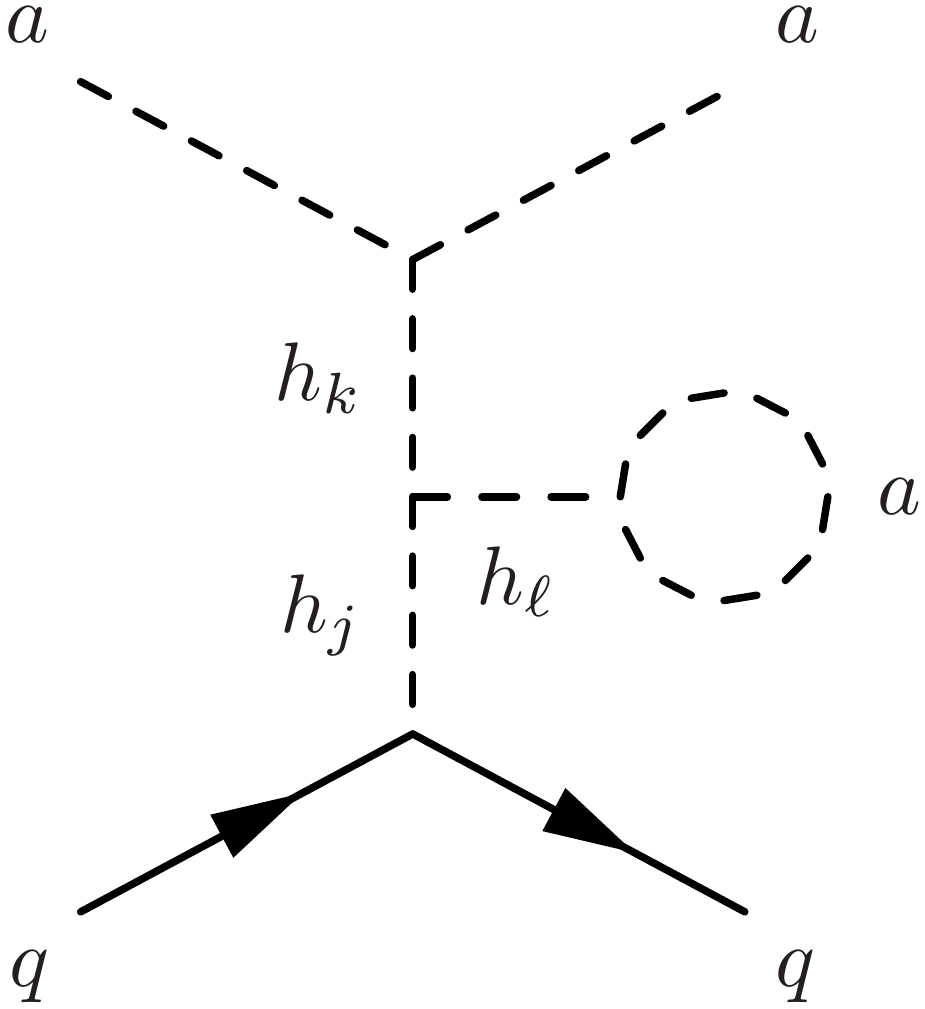}
\includegraphics[width=0.25\textwidth]{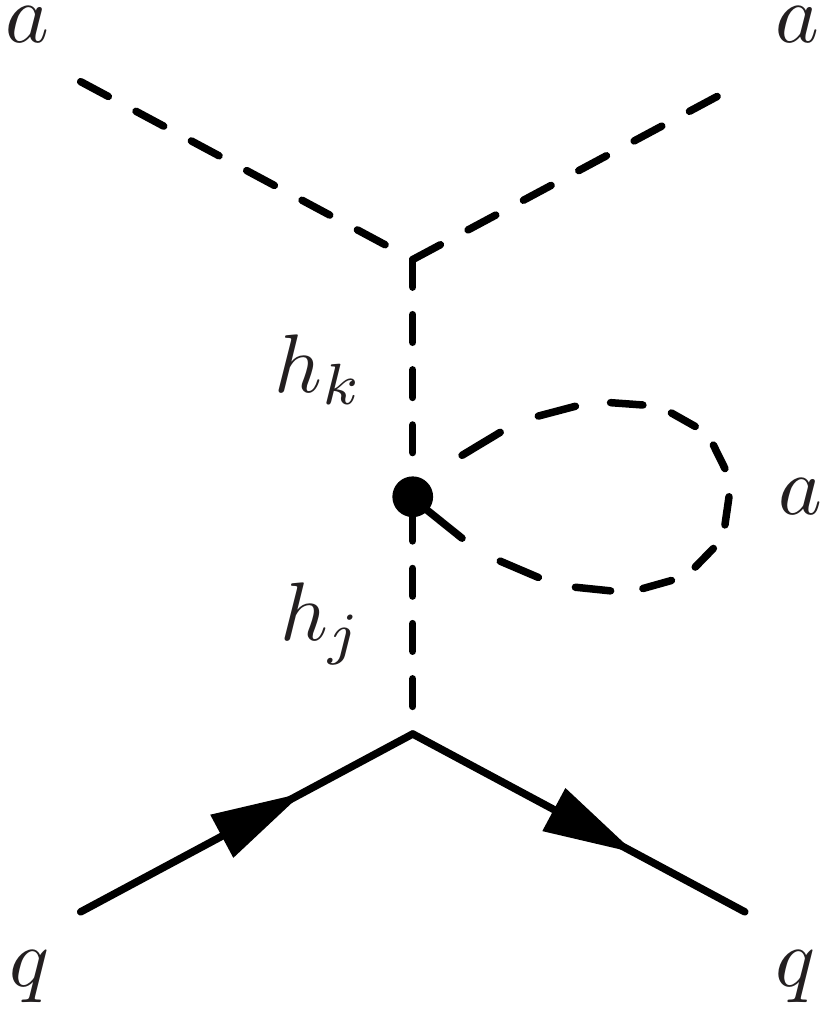}
\caption{
The diagrams that cancel each other.
}
\label{fig:ningen}
\end{figure}

\begin{figure}[tbp]
\centering
\subfigure[]{\includegraphics[width=0.25\textwidth]{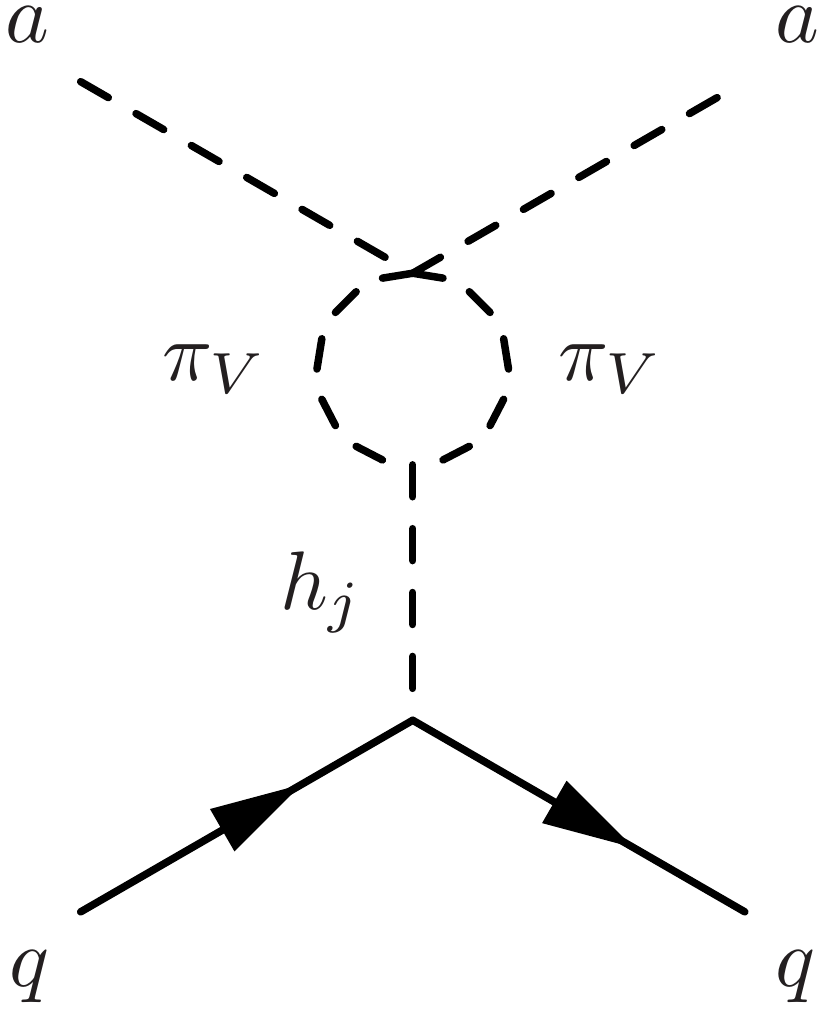}\label{subfig:a_kkD}}
\subfigure[]{\includegraphics[width=0.25\textwidth]{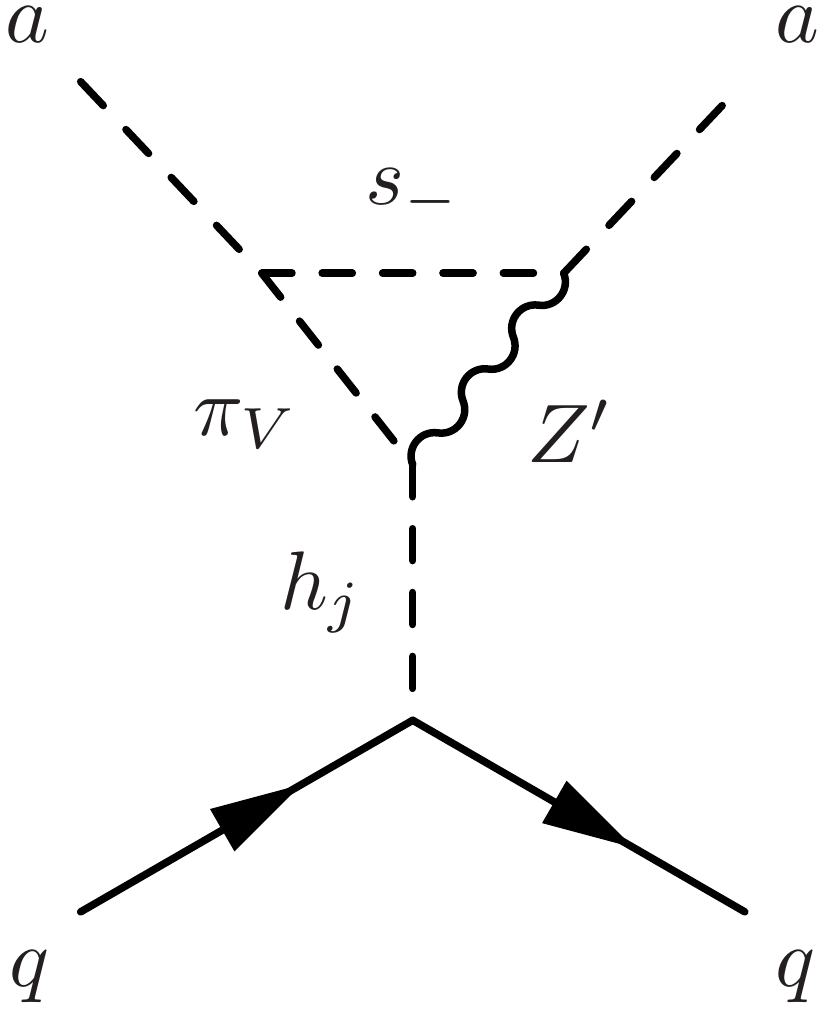}}
\subfigure[]{\includegraphics[width=0.25\textwidth]{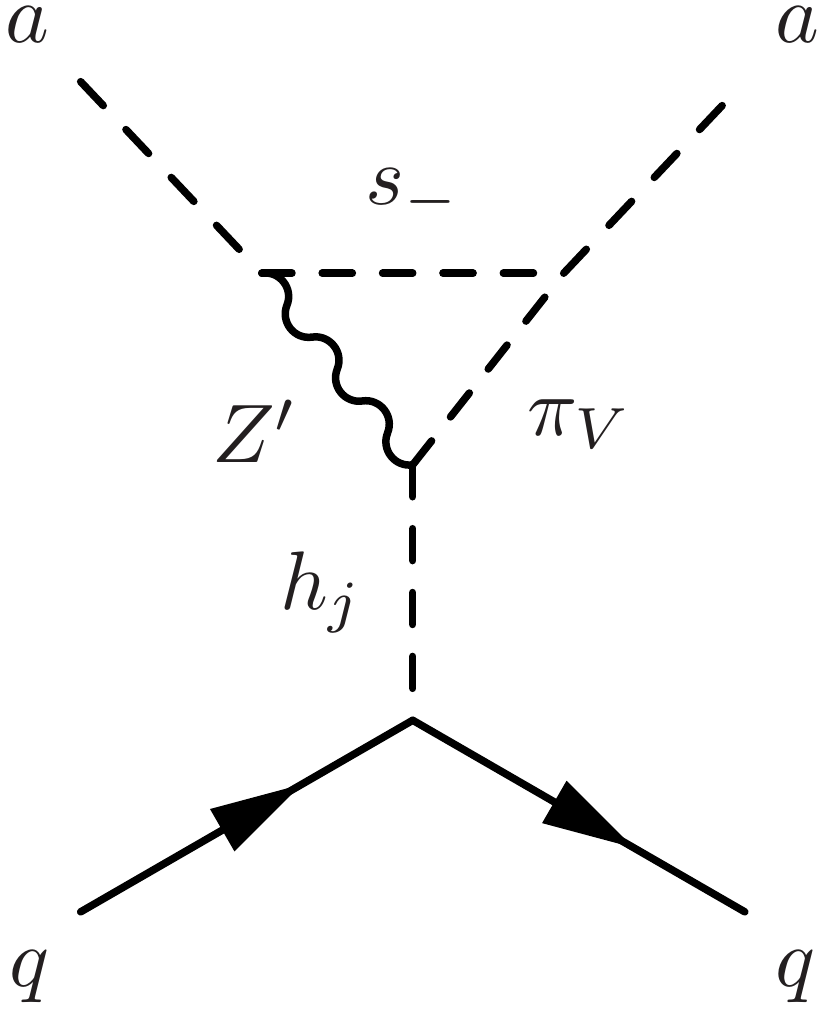}}
\caption{
This diagram does not contribute to $\sigma_\text{SI}^{}$. See text.
}
\label{fig:landau0}
\end{figure}

\begin{figure}[tbp]
\centering
\subfigure[]{\includegraphics[width=0.25\textwidth]{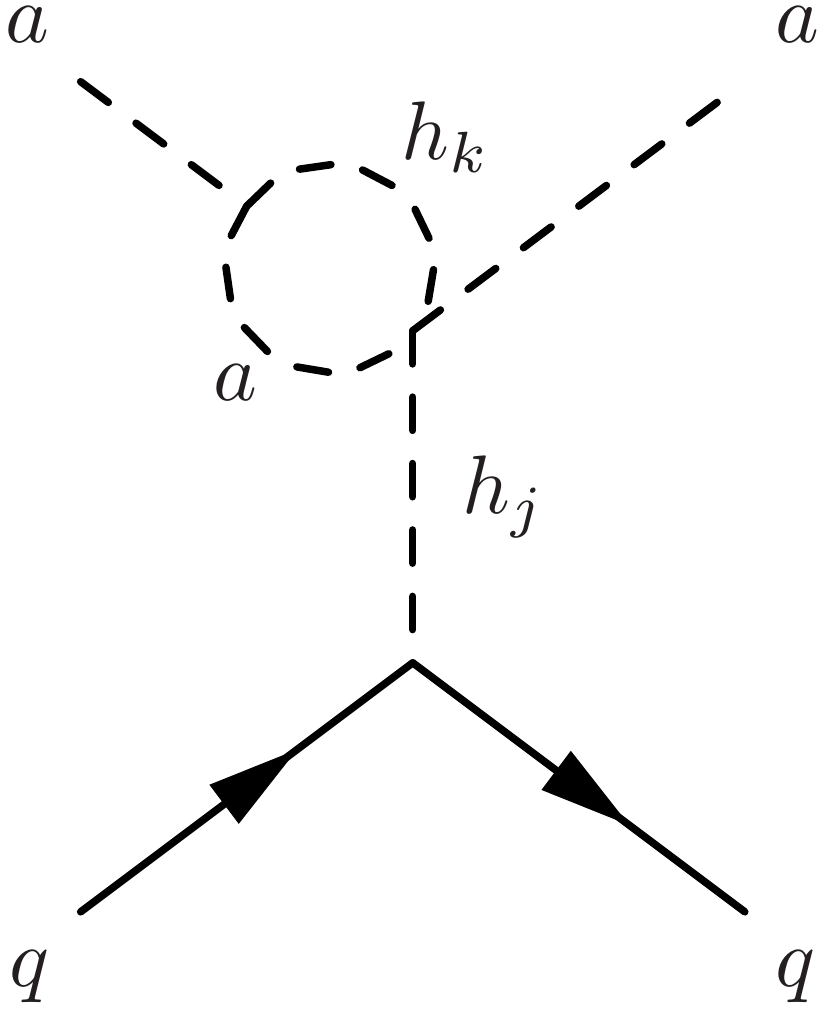}\label{subfig:a_kkL}}
\ \
\subfigure[]{\includegraphics[width=0.25\textwidth]{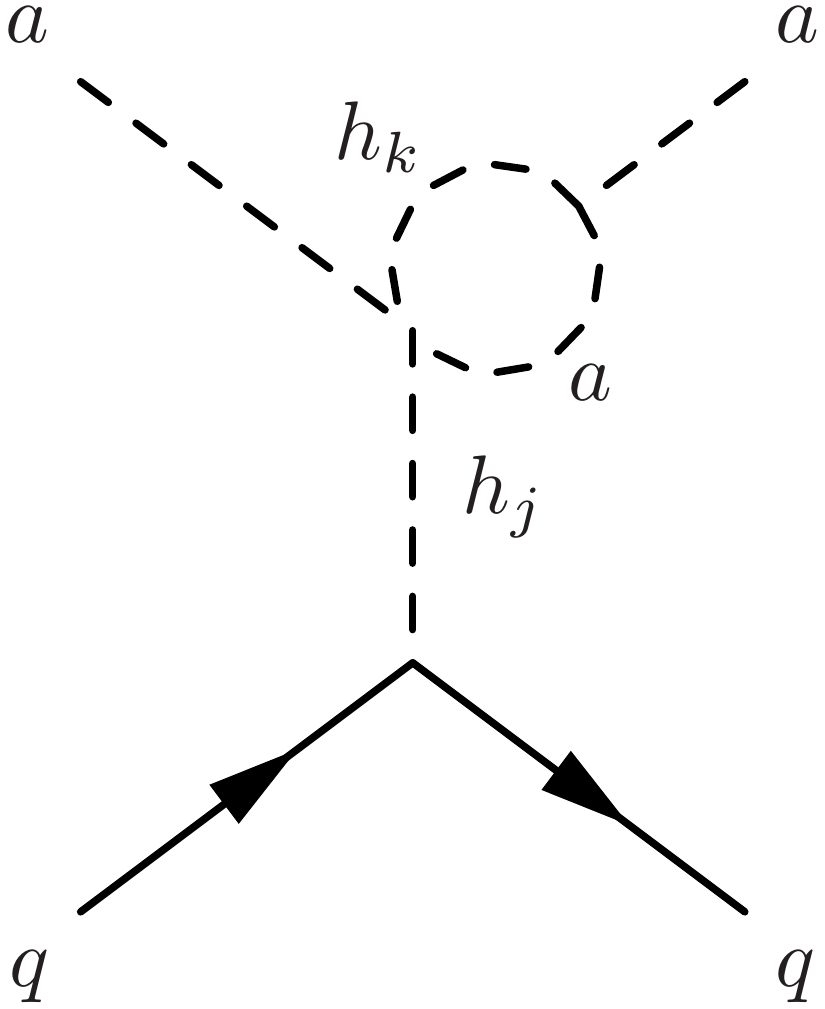}\label{subfig:a_kkR}}
\ \
\subfigure[]{\includegraphics[width=0.25\textwidth]{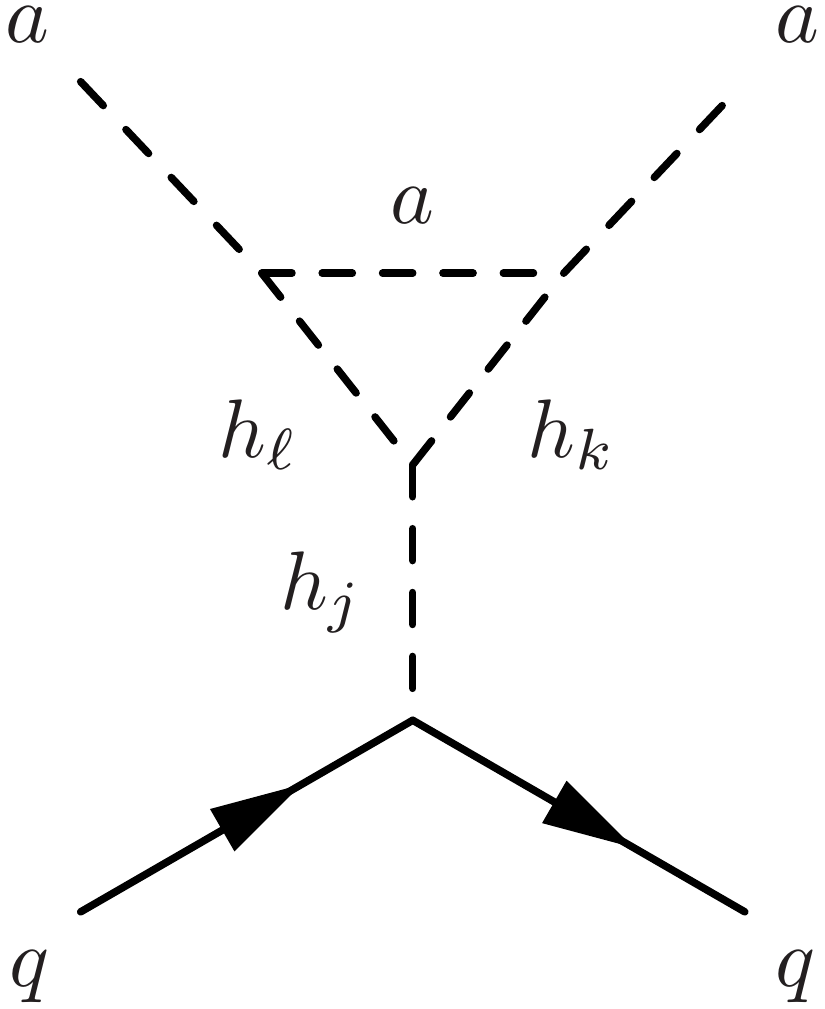}\label{subfig:a_vertex}}
\caption{
The diagrams that contain $a$ in the loop.
}
\label{fig:diagram_with_a}
\end{figure}

\begin{figure}[tbp]
\centering
\subfigure[]{\includegraphics[width=0.23\textwidth]{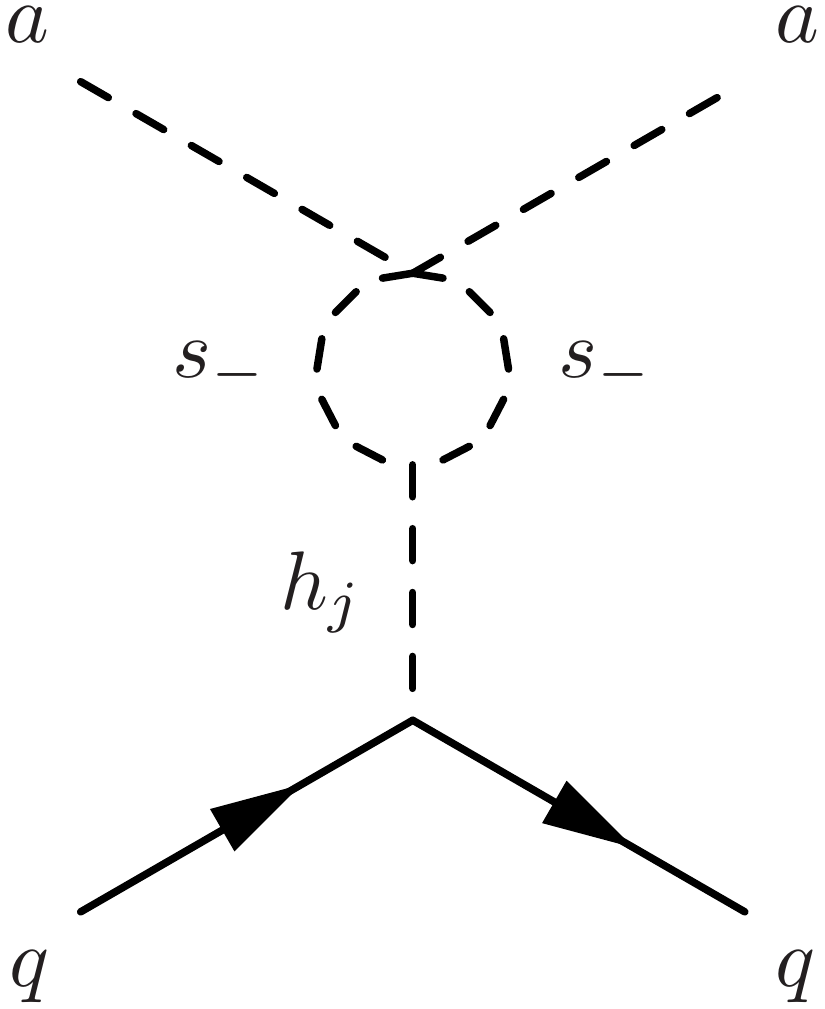}\label{subfig:s_0}}
\ \
\subfigure[]{\includegraphics[width=0.23\textwidth]{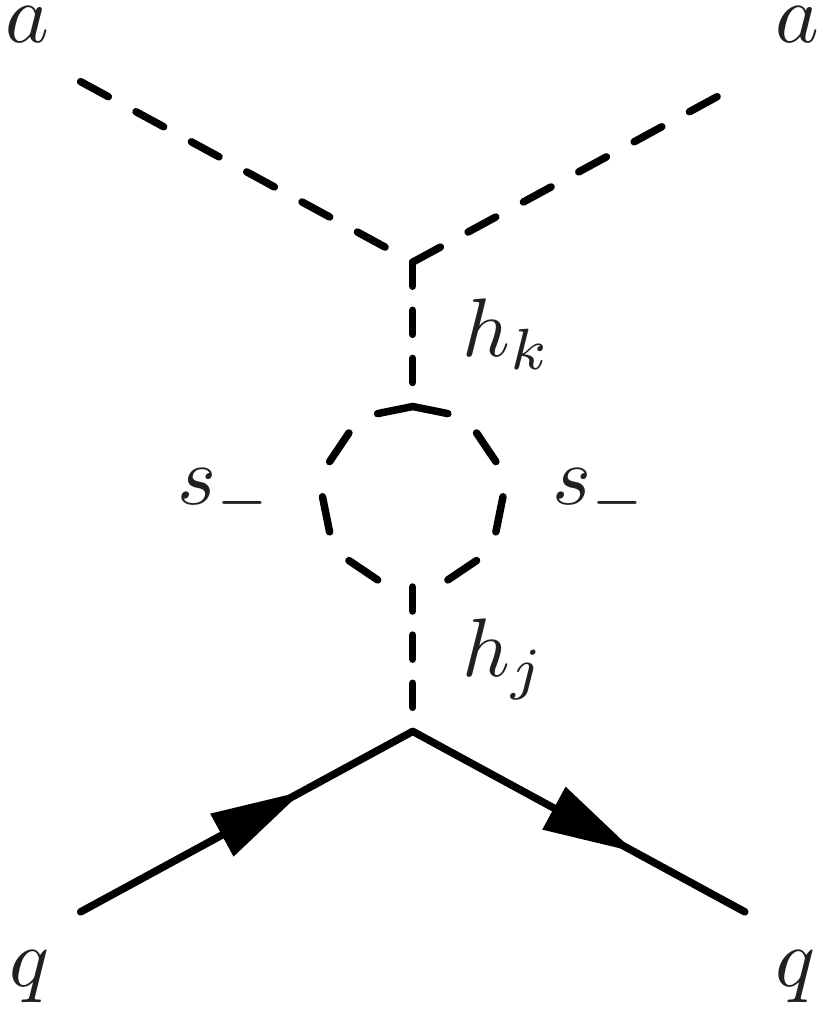}\label{subfig:s_1}}
\ \
\subfigure[]{\includegraphics[width=0.23\textwidth]{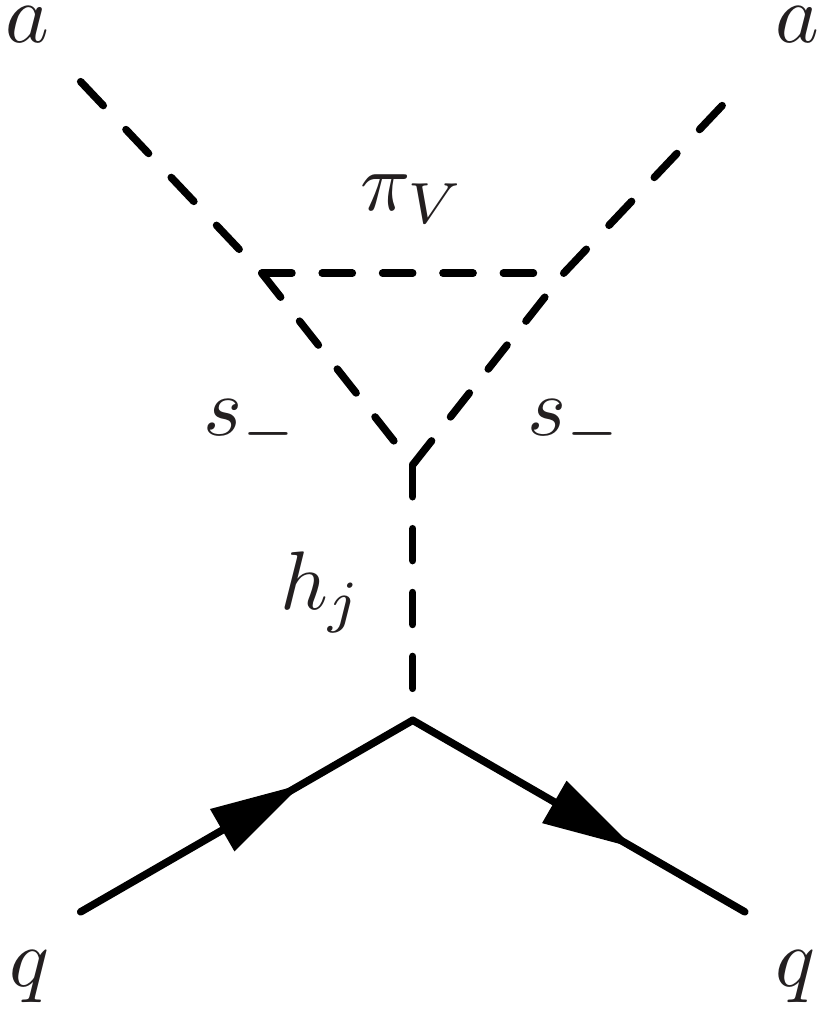}\label{subfig:s_2}}
\ \
\subfigure[]{\includegraphics[width=0.23\textwidth]{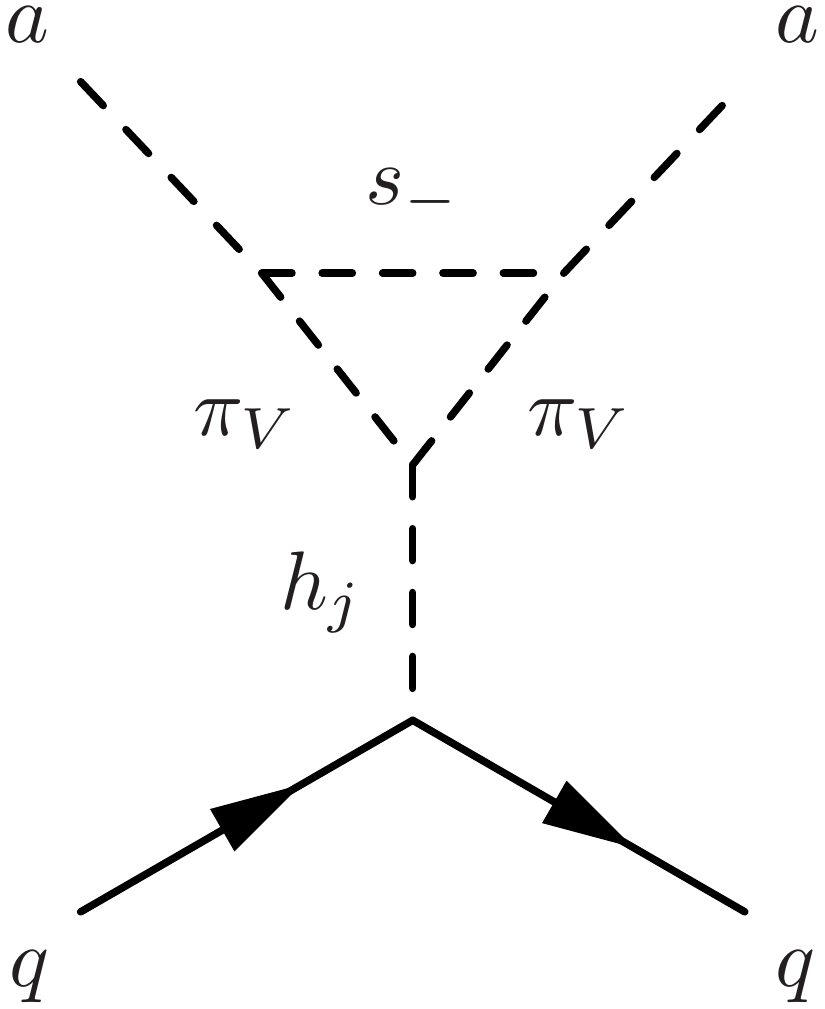}\label{subfig:s_3}}
\ \\ \ \\ 
\subfigure[]{\includegraphics[width=0.23\textwidth]{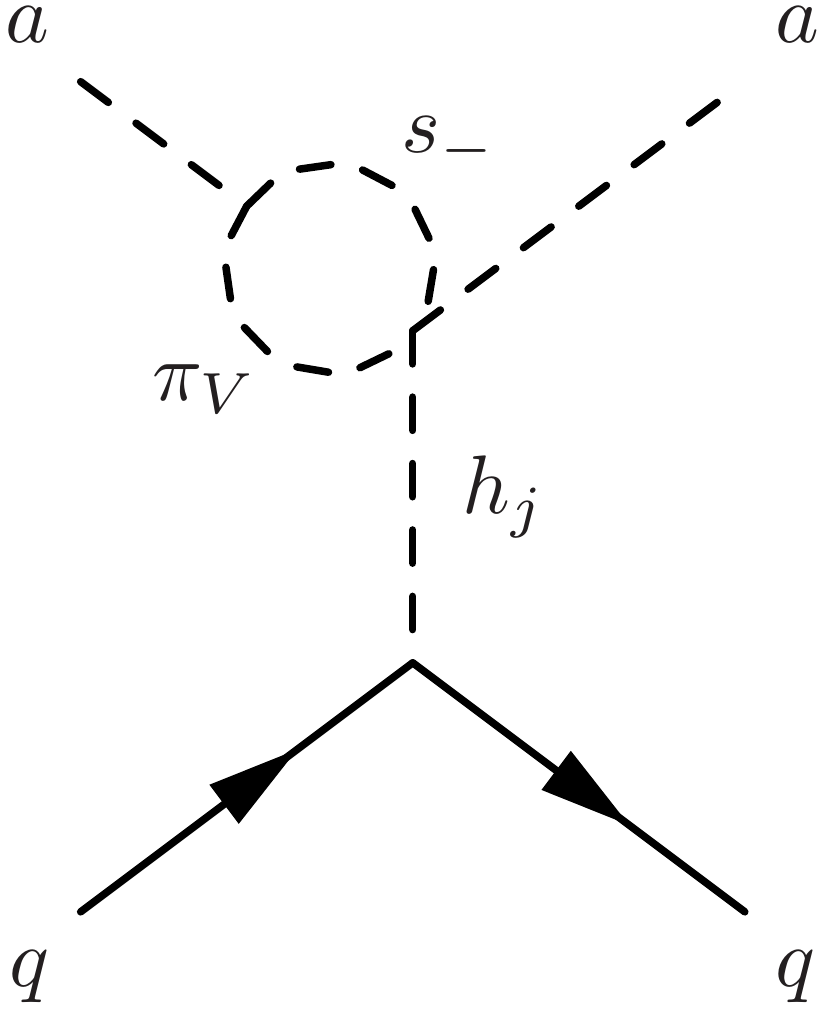}\label{subfig:s_4}}
\ \
\subfigure[]{\includegraphics[width=0.23\textwidth]{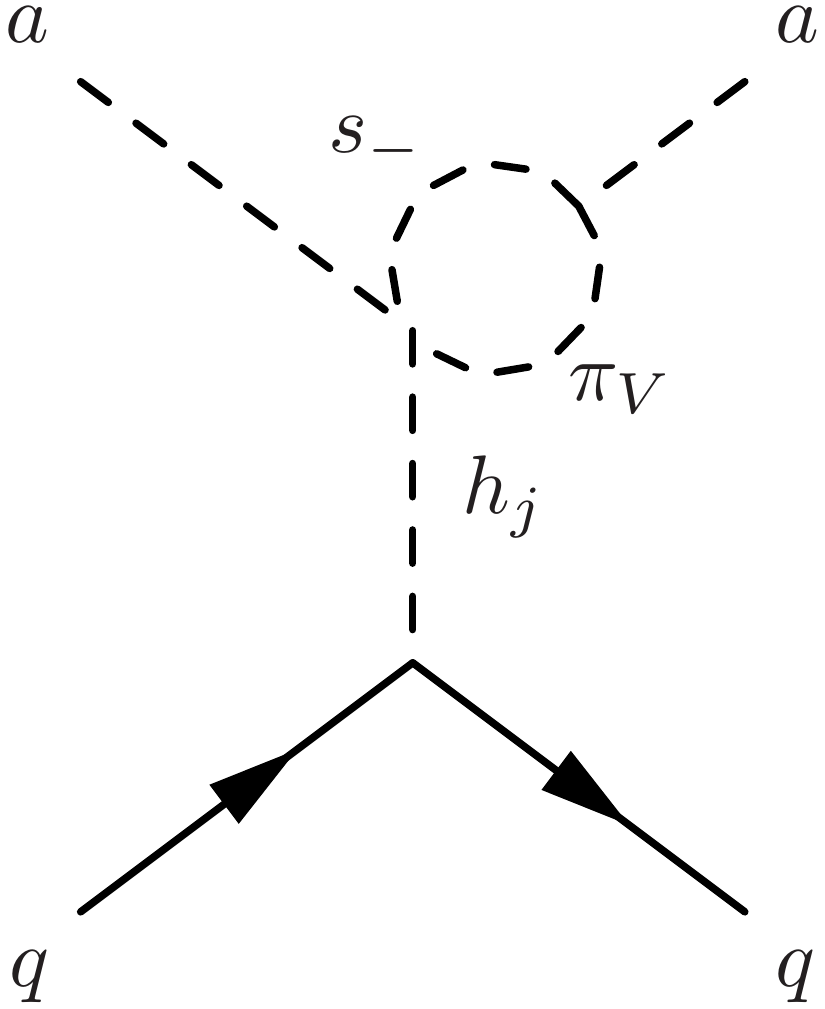}\label{subfig:s_5}}
\ \
\subfigure[]{\includegraphics[width=0.23\textwidth]{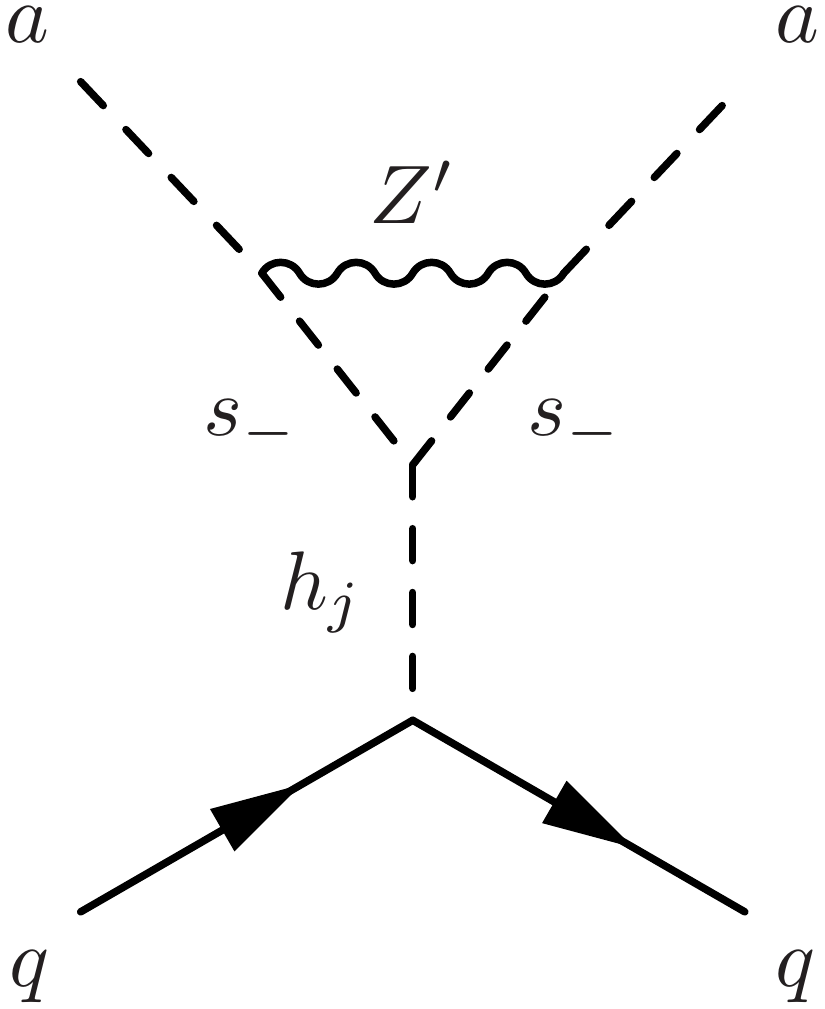}\label{subfig:s_V1}}
\ \
\subfigure[]{\includegraphics[width=0.23\textwidth]{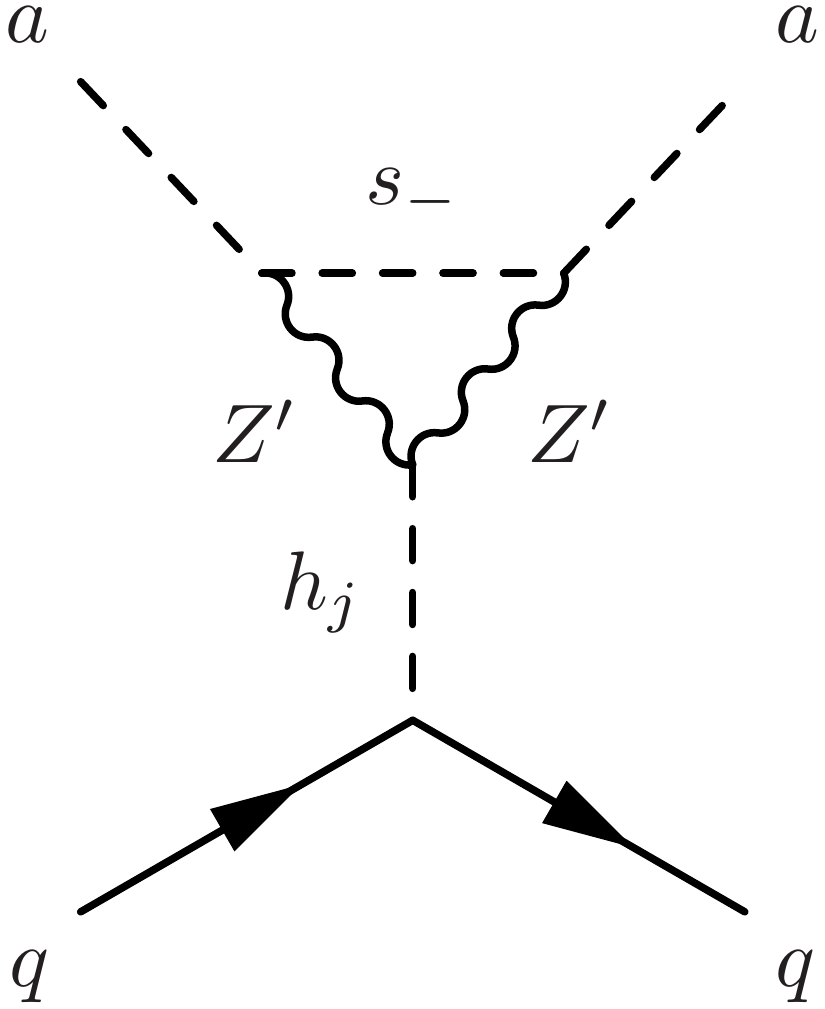}\label{subfig:s_V2}}
\caption{
The diagrams that contain $s$ in the loop.
}
\label{fig:diagram_with_s}
\end{figure}

\begin{figure}[tbp]
\centering
\includegraphics[width=0.23\textwidth]{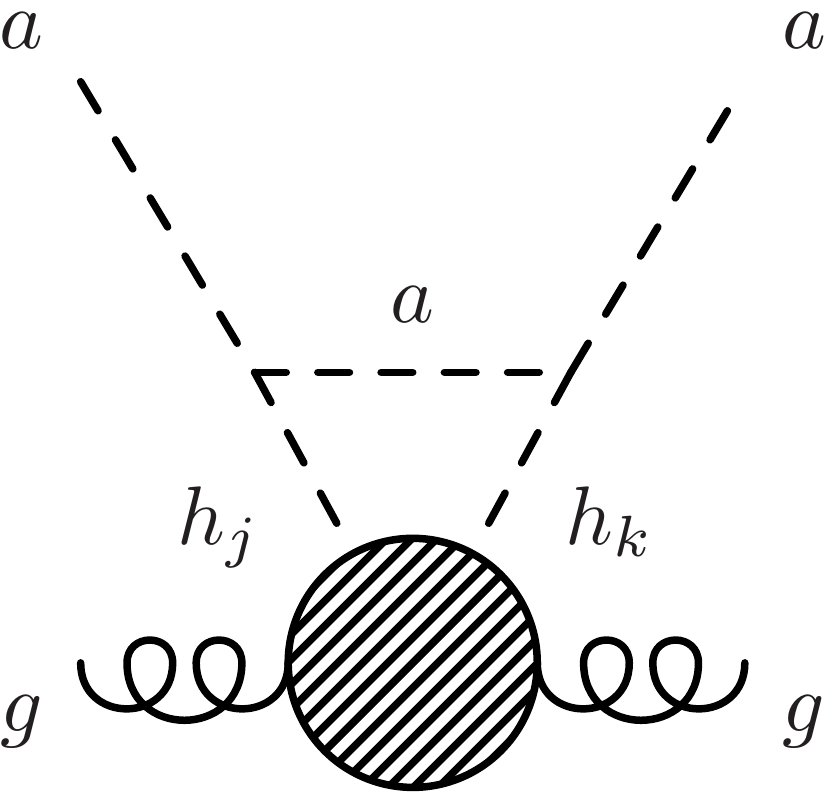}
\caption{
The 2-loop diagram that contains top loop in the blob.
}
\label{fig:2-loop}
\end{figure}

As discussed in Sec.~\ref{sec:loop},
we focus only on the diagrams that depend on the parameter $m_{12}^2$, explicit breaking of the global $U(1)_a$ symmetry.
The $m_{12}^2$ dependence appears in $m_{a}$, $m_{s_{-}}$, the vertices including $s_{-}$, and the $a a \pi_{V}^{} \pi_{V}^{}$-coupling, where $\pi_V^{}$ is the would-be NG boson that is eaten by $Z'$.
We list the diagrams in Figs.~\ref{fig:ningen}--
\ref{fig:2-loop}.

Some of these diagrams do not contribute to our final result.
In fact, the diagrams in Fig.~\ref{fig:ningen} cancel each other.
The diagrams in Fig.~\ref{fig:landau0} depend on the gauge choice, but they vanish in the Landau gauge with the dimensional regularization in the limit of the zero momentum transfer. Also, $\pi_V^{} \pi_V^{} h_j$ coupling is proportional to $aah_j$ coupling, and thus Fig.~\ref{subfig:a_kkD} vanishes even with other gauge choices. 
Hence, those shown in Figs.~\ref{fig:ningen} and \ref{fig:landau0} do not contribute to $\sigma_\text{SI}^{}$.

Figure \ref{fig:diagram_with_a} shows the diagrams containing $a$ in the loop.
We find
\begin{align}
\text{Fig.~\ref{subfig:a_kkL}}
=&~\text{Fig.~\ref{subfig:a_kkR}}
= -\frac{i}{(4\pi)^2}\bar{u}u \sum_{j,k} g_{ffh_j} \frac{1}{m_{h_j}^2} g_{aah_j h_k} g_{aah_k}B_0(m_a^2,m_a^2, m_{h_j}^2)
,\\
\text{Fig.~\ref{subfig:a_vertex}}
=& -\frac{i}{(4\pi)^2}\bar{u}u \sum_{j,k,\ell} g_{ffh_j} \frac{1}{m_{h_j}^2} g_{h_j h_k h_\ell} g_{aah_\ell} 
\frac{B_0(m_a^2,m_a^2, m_{h_k}^2)-B_0(m_a^2,m_a^2, m_{h_\ell}^2)}{m_{h_k}^2-m_{h_\ell}^2}
,
\end{align}
where
\allowdisplaybreaks[1]
\begin{align}
g_{ffh}=& \frac{m_f}{v} c_\xi,\\ 
g_{ffh'}=& -\frac{m_f}{v} s_\xi,\\
g_{aah}=& \frac{m_h^2}{v_s} s_\xi,\\ 
g_{aah'}=& \frac{m_{h'}^2}{v_s} c_\xi,\\ 
g_{aahh}=& \frac{(m_{h'}^2 c_\xi^2  s_\xi  + m_h^2 s_\xi^3 )v + c_\xi^3 (m_h^2 - m_{h'}^2) v_s}{v v_s^2} s_\xi,\\ 
g_{aahh'}=& \frac{(m_{h'}^2 c_\xi^2   + m_h^2 s_\xi^2) v - c_\xi s_\xi (m_h^2 - m_{h'}^2) v_s}{v v_s^2} s_\xi c_\xi,
\\ 
g_{aah'h'}=& \frac{(m_{h'}^2 c_\xi^3   + m_h^2 c_\xi s_\xi^2 )v + s_\xi^3 (m_h^2 - m_{h'}^2) v_s}{v v_s^2} c_\xi,
\\ 
g_{hhh}=& \frac{3 m_h^2}{v} \frac{s_\xi^3 v + c_\xi^3 v_s }{v_s},
\\ 
g_{hhh'}=& \frac{2 m_h^2 + m_{h'}^2}{v v_s} s_\xi c_\xi (s_\xi v -c_\xi v_s),
\\ 
g_{hh'h'}=& \frac{m_h^2 + 2 m_{h'}^2}{v v_s} s_\xi c_\xi (c_\xi v +s_\xi v_s),
\\ 
g_{h'h'h'}=& \frac{3 m_{h'}^2}{v} \frac{c_\xi^3 v - s_\xi^3 v_s }{v_s},
\end{align}
and the definitions of the loop functions are the same as those given by \texttt{Looptools}~\cite{Hahn:1998yk}.
Note that 
the contributions from Fig.~\ref{fig:diagram_with_a} do not vanish in the limit of $m_{12}^2 \to 0$
unless adding other diagrams that are independent of $m_{12}^2$.
Instead of adding them, we subtract $m_{12}^2$ independent part as stated in Sec.~\ref{sec:loop},

Figure~\ref{fig:diagram_with_s} shows the diagrams containing $s_-$ in the loop.
Among these diagrams, Fig.~\ref{subfig:s_3} vanishes because $\pi_V^{} \pi_V^{} h_j$ coupling is proportional to $aah_j$ coupling.
We calculate the rest of the diagrams. We find
\allowdisplaybreaks[1]
\begin{align}
\text{Fig.~\ref{subfig:s_0}}
+ \text{Fig.~\ref{subfig:s_1}}
=&
\frac{i}{(4\pi)^2} \bar{u}u \frac{m_q}{v}
2 \frac{(m_a^2-m_{s_-}^2)^2}{v_s^3} s_\xi c_\xi \left(\frac{1}{m_h^2}-\frac{1}{m_{h'}^2}\right)
B_0(0,m_{s_-}^2,m_{s_-}^2)
,\\
\text{Fig.~\ref{subfig:s_2}}
=&
\frac{i}{(4\pi)^2} \bar{u}u \frac{m_q}{v}\frac{(m_a^2-m_{s_-}^2)^3}{v_s^3} 
\left(\frac{1}{m_h^2}-\frac{1}{m_{h'}^2}\right)
2 s_\xi c_\xi 
\pdv{m_{s_-}^2} B_0(m_a^2, 0, m_{s_-}^2)
,\\
\text{Fig.~\ref{subfig:s_3}}
=&
0
,\\
\text{Fig.~\ref{subfig:s_4}}
+\text{Fig.~\ref{subfig:s_5}}
=&
-\frac{i}{(4\pi)^2} \bar{u}u \frac{m_q}{v}\frac{(m_a^2-m_{s_-}^2)^2}{v_s^3} 
\left(\frac{1}{m_h^2}-\frac{1}{m_{h'}^2}\right)
2 s_\xi c_\xi 
B_0(m_a^2, 0, m_{s_-}^2)
,\\
\text{Fig.~\ref{subfig:s_V1}}
=&
-\frac{i}{(4\pi)^2} \frac{m_q}{v}\bar{u}u 
 s_\xi c_\xi \frac{2 m_{Z'}^2 (m_a^2-m_{s-}^2)}{v_s^3}
\left(\frac{1}{m_h^2}-\frac{1}{m_{h'}^2}\right)
\nonumber\\
&
\times \Biggl\{
4 m_a^2 \pdv{m_{s-}^2} B_0(m_a^2, m_{Z'}^2, m_{s_-}^2)
\nonumber\\
& \qquad 
-\frac{(m_a^2-m_{s_-}^2)^2}{m_V^2}
\pdv{m_{s_-}^2}
\left(
B_0(m_a^2, m_{Z'}^2,m_{s_-}^2) - B_0(m_a^2, 0,m_{s_-}^2)
\right)
\nonumber\\
& \qquad 
+\frac{2(m_a^2-m_{s_-}^2)}{m_V^2}
\left(
B_0(m_a^2, m_{Z'}^2,m_{s_-}^2) - B_0(m_a^2, 0,m_{s_-}^2)
\right)
\nonumber\\
& \qquad 
-( 2(m_a^2-m_{s_-}^2)+m_V^2)
\pdv{m_{s_-}^2} B_0(m_a^2, m_{Z'}^2,m_{s_-}^2)
\nonumber\\
& \qquad 
-\frac{1}{m_V^2} A_0(m_V^2)
+ 2 B_0(m_a^2, m_{Z'}^2,m_{s_-}^2)
- B_0(0, m_{s_-}^2,m_{s_-}^2)
\Biggr\}
,\\
\text{Fig.~\ref{subfig:s_V2}}
=&
-2\frac{i}{(4\pi)^2} \frac{m_q}{v}\bar{u}u 
  s_\xi c_\xi \frac{m_{Z'}^4}{v_s^3}
\left(\frac{1}{m_h^2}-\frac{1}{m_{h'}^2}\right)
\nonumber\\
&
\times \Biggl\{
\left( -m_{Z'}^2 + 2 m_{s_-}^2 + 2 m_a^2 \right)
\pdv{m_{Z'}^2} B_0(m_a^2, m_{Z'}^2, m_{s_-}^2)
\nonumber\\
& \qquad 
 +B_0(0, m_{Z'}^2, m_{Z'}^2)
 - B_0(m_{a}^2, m_{Z'}^2, m_{Z'}^2)
\nonumber\\
& \qquad 
-\frac{(m_{s_-}^2 - m_a^2)^2}{m_V^2} 
\pdv{m_{Z'}^2} B_0(m_a^2, m_{Z'}^2, m_{s_-}^2)
\nonumber\\
& \qquad 
 +\frac{(m_{s_-}^2 - m_a^2)^2}{m_V^4} 
 \left( 
 B_0(m_a^2, m_{Z'}^2, m_{s_-}^2) - B_0(m_a^2, 0, m_{s_-}^2)
 \right)
\nonumber\\
& \qquad 
+ \frac{m_{s_-}^2-m_a^2}{m_V^2}
\Biggr\}
.
\end{align}
Note that the amplitude from Fig.~\ref{fig:diagram_with_s} vanishes in $m_{12}^2 \to 0$ limit. 
Thus we have
\begin{equation}
\mathcal{M}_\text{Fig.~\ref{fig:diagram_with_s}}(m_{12}^2) 
-\mathcal{M}_\text{Fig.~\ref{fig:diagram_with_s}}(0)
=\mathcal{M}_\text{Fig.~\ref{fig:diagram_with_s}}(m_{12}^2) \, .
\end{equation}

One should also consider the gluon contribution with the heavy quarks in loop diagrams shown in  
Fig.~\ref{fig:2-loop}.
Although it is the two-loop contribution, it contributes as much as the one-loop diagrams discussed so far. 
The expression after subtracting $m_{12}^2$ dependence is given in Ref.~\cite{Abe:2022mlc}.

To obtain $\sigma_\mr{SI}^{}$,
it is convenient to consider the effective Lagrangian relevant for the scattering process as given by
\begin{equation}
 \mathcal{L}_{eff} = \frac{1}{2} C_q^S \, m_q a^2 \bar{q} q + \frac{1}{2}C_g^S \, \frac{\alpha_s}{\pi} a^2 G^a_{\mu\nu} G^{a \mu\nu},
 \end{equation}
where $C_q^S $ and $C_g^S $ are Wilson coefficients
and can be read off from the scattering amplitudes calculated so far.
Using these Wilson coefficients, we calculate $\sigma_\text{SI}^{}$,
\begin{equation}
    \sigma_\mathrm{SI}^{} = \frac{1}{4\pi} \left( \frac{\mu_N m_N}{m_a}\right)^2 
\left| \sum_{q=u,d,s} C_q^S f^N_q - \frac{8}{9} C_g^S f_g^N\right|^2 ,
\end{equation}
where $m_N$ and $\mu_N\equiv m_N m_a/(m_N + m_a)$ are the nucleon mass and the reduced mass, respectively.
$f^N_q$ and $f_g^N$ are the matrix elements of the operators evaluated by the nucleon states,
\begin{align}
 f^N_q m_N &= \bra{N}m_q \bar q q \ket{N} , \\
 -\frac{8}{9}f^N_g m_N &= \bra{N}\frac{\alpha_s}{\pi} G^a_{\mu\nu}G^{a \mu\nu} \ket{N}.
\end{align}
Their approximate values are given as \cite{Hisano:2015rsa}
\begin{equation}
 f_u^p = 0.019, \quad f_d^p = 0.027, \quad f_u^n = 0.013, \quad f_d^n = 0.040, \quad  f_s^p=f_s^n = 0.009,
\end{equation}
\begin{equation}
 f_g^N  = 1-f_u^N-f_d^N-f_s^N + \mathcal{O} (\alpha_s).
\end{equation}

\bibliographystyle{jhep}
\bibliography{./references}

\end{document}